%% file: phast.tex
\documentclass[twoside,leqno,twocolumn]{article}

\usepackage[letterpaper]{geometry}

\usepackage{siamproceedings}

\usepackage[T1]{fontenc}
\usepackage{amsfonts}
\usepackage{graphicx}
\usepackage{epstopdf}
\usepackage{enumitem}
\usepackage{algorithmic}
\ifpdf
  \DeclareGraphicsExtensions{.eps,.pdf,.png,.jpg}
\else
  \DeclareGraphicsExtensions{.eps}
\fi

\usepackage{amsopn}

\usepackage{tikz}
\usetikzlibrary{matrix, positioning, decorations.pathreplacing, arrows, arrows.meta, calc, fit}
\usepackage{pgfplots}
\usepackage{pgfplotstable}
\usepackage{multirow}
\usepackage{booktabs}

\newcommand{\seeds}{\mathit{seeds}} 
\DeclareMathOperator{\hi}{hi}
\DeclareMathOperator{\lo}{lo}

\DeclareMathOperator{\round}{round}
\newcommand{\fmap}[2]{\left\lfloor \frac{#2}{2^u} \cdot #1 \right\rfloor}
\newcommand{\ps}[1]{}
\newcommand{\pb}[1]{}
\newcommand{\PHastPlus}{PHast$^+$}
\begin{document}

\title{PHast -- \underline{P}erfect \underline{H}ashing made f\underline{ast}}
\author{Piotr Beling\thanks{Faculty of Mathematics and Computer Science, University of Łódź, Poland (\email{piotr.beling@wmii.uni.lodz.pl}, \url{http://pbeling.w8.pl/}, \url{https://orcid.org/0000-0003-3048-3704}).}
\and
Peter Sanders\thanks{Karlsruhe Institute of Technology, Germany (\email{sanders@kit.edu}, \url{https://orcid.org/0000-0003-3330-9349}).}
}


\date{}


\maketitle

\begin{abstract}
Perfect hash functions give unique ``names'' to arbitrary keys requiring only a few bits per key. This is an essential building block in applications like static hash tables, databases, or bioinformatics. This paper introduces the PHast approach that combines the fastest available queries, very fast construction, and good space consumption (below 2 bits per key).
PHast improves \emph{bucket-placement} which first hashes each key $k$ to a bucket, and then looks for the bucket seed $s$ such that a placement function maps pairs $(s,k)$ in a collision-free way. PHast can use small-range hash functions with linear mapping, fixed-width encoding of seeds, and parallel construction.
This is achieved using small overlapping slices of allowed values and bumping to handle unsuccessful seed assignment.
A variant we called \PHastPlus{} uses additive placement, which enables bit-parallel seed searching, speeding up the construction by an order of magnitude.
\end{abstract}

\textbf{Source code available at:} \\ 
\url{https://doi.org/10.6084/m9.figshare.30272380.v6}

\input{introduction}
\input{method}

\input{details}

\input{plus}
\input{parallel}
\input{benchmarks}
\input{external_memory}
\input{advanced}

\input{conclusions}


\bibliographystyle{siamplain}
\bibliography{phast}


\end{document}

%% file: introduction.tex
\section{Introduction.}\label{sec:introduction}

We call $f:\{k_1,\ldots,k_n\}\rightarrow \{0,\ldots,m-1\}$ a \emph{perfect hash function (PHF)} if
$f$ is injective. We call it \emph{minimal perfect} (MPHF) if, in addition,  $m=n$. (M)PHFs have been intensively studied as they are an important building block for databases, text indexing, and some bioinformatics applications. 
When designing MPHFs, we face a multicriteria optimization problem because we want to simultaneously optimize space consumption, construction time, and query time. There is a lower bound of $\log_2e\approx 1.44$ bits per key \cite{MPHFLowerBound1984, MPHFLowerBound1992} of an MPHF. Recent work hat put particular emphasis on practical constructions that approach this lower bound.

This paper departs from a different starting point, asking what is the fastest possible query time for a PHF.
Surprisingly, we obtain a scheme with not only very fast evaluation, but also construction. This is particularly relevant in practice, as PHFs are often used to accelerate static data structures.

PHast (Perfect Hashing made fast) is based on the particularly successful approach of \emph{bucket placement} \cite{BelazzouguiBD09,pibiri2021pthash,HLPSW24,grootkoerkamp:LIPIcs.SEA.2025.21}:
A key $k$ is first hashed to a $u$-bit hash code $h(k)$.
As a next step, a \emph{bucket assignment function} $\gamma$ maps $h(k)$ to a bucket index $i$.
A \emph{seed} $s=\seeds[i]$ is then retrieved.
Finally, a placement function $p$ computes the actual function value $p(s, h(k)) \in \{0,\ldots,m-1\}$.

In order for this to work with fast construction and low storage, previous schemes incur additional query time by requiring a nonlinear $\gamma$
and/or variable bit length encoding for the seed array.
They also need $u>2\log_2 n$ and a way to seed the hash function so that
its values do not collide.

A contribution of PHast is to show that all this can be massively simplified.
It suffices that $h$ is fixed and produces $u=\log_2 n + O(1)$ bits that are allowed to collide.
PHast uses a linear bucket assignment function and fixed-width seeds, that is $\seeds$ is an array of $S$-bit integers (for example $S=8$)
and the used bucket index is simply $i=\lfloor \beta  h(k)\rfloor$.
This is achieved by two small deviations from the general scheme described above.
First, we add a linear offset $\lfloor \alpha h(k) \rfloor$ and restrict the range of $p$ to $\{0,\ldots,L-1\}$, for example with $L=1024$.
As a~welcome side effect, this also improves the cache-efficiency of construction.
Second, buckets for which no seed can be found are bumped. Bumping is indicated with the reserved seed value $0$. Keys in bumped buckets are handled in a special, more expensive way. However, as only few buckets (typically about $1\%$) are bumped, the expected query time is hardly affected.
Construction can be further accelerated by using an additive placement function.
The search for an appropriate seed can then be done in a bit-parallel way by computing the bit-or of occupied positions.

To summarize, a~key $k$ is first hashed using a hash function
$h$. Then, the seed $s=\seeds[\lfloor\beta h(k)\rfloor]$ is read.
The hash function value is then
\begin{equation}\label{eq:query}
f(k)=\begin{cases}\lfloor \alpha h(k) \rfloor + p(s, h(k))& \text{ if $s>0$}\\
\text{fallback\_for\_bumped}(k) & \text{ else}\\
  \end{cases}
  \end{equation}
where $\alpha$ and $\beta$ are scaling factors.

The construction algorithm orders the buckets based on their index and size.
This suffices to make almost all buckets similarly easy to place and thus obviates the need for a nonlinear bucket assignment function or variable bit length encoding. 

We were able to implement it effectively thanks to two innovative solutions:
\begin{itemize}
    \item Seeds are assigned to buckets in approximate order from smallest, to largest indexes. As a result, function values are also taken up approximately from smallest to largest. 
    These facts enable the implementation to be CPU-cache and parallelization friendly.
    
    \item Unlike other methods, PHast does not assign the first feasible seed found for a bucket. Instead, it picks the one that occupies the smallest possible function values, thus heuristically maximizing the chance of successful seeding further buckets.
\end{itemize}
We continue this paper with a  review of related work in Section~\ref{sec:related}.
Then Section~\ref{sec:phast} introduces PHast, while Section \ref{sec:parallel} describes the parallel algorithm for its construction.
Section~\ref{sec:benchmark} gives an extensive experimental evaluation.
In Section~\ref{sec:conclusions} we summarize the results and discuss future directions for research.
Section \ref{ss:external} descries constructing PHast in external memory, while Section \ref{s:advanced} summarizes our attempts to reduce PHast size via some advanced techniques.

\paragraph*{Summary of Contributions:}
\begin{itemize}
\item By linearly mapping buckets to small, overlapping slices, we avoid the nonlinear bucket assignment and enable cache-efficient and multithreaded construction. In contrast to partitioning approaches used elsewhere, this multithreaded construction incurs no query overhead.
\item Bumping allows PHast to use a small-range (and thus fast) hash function and to avoid variable bit length encoding or complicated local search that requires considerable more space to work.
\item Heuristic seed assignment reduces space consumption.
\item A very simple (bit-parallel) construction algorithm.
\item Extensive experimental evaluation.
\item Beyond perfect hashing, PHast can be considered an interesting case study in how to engineer a data structure for speed and simplicity.
\end{itemize}





\section{Related Work.}\label{sec:related}
A lot of work has been done on perfect hashing.
See \cite{lehmann2025modernminimalperfecthashing} for a recent overview.

\emph{Fingerprinting} approaches \cite{chapman2011meraculous,muller2014retrieval,limasset2017fast,beling2023fingerprinting,Lehmann24a} allow very fast construction
but need at least $e$ bits per key (even more when
fast evaluation is the goal). Even in the best case, a query requires a rank operation in a bit vector which is slower than a PHast query.

Approaches based on \emph{static function retrieval} \cite{chazelle2004bloomier,botelho2007simple,genuzio2016fast,lehmann2023sichash}
encode a per-key seed. However, this requires longer query time
than just using a per-bucket seed. The reason is that the retrieval operation either
probes multiple memory positions or requires complex calculations based on several contiguous machine words \cite{dillinger2022burr}. 

The currently most space efficient approaches to perfect hashing use \emph{brute-force search for PHFs of small subproblems} \cite{esposito2020recsplit,bez2023high,lehmann2023shockhash,lehmann2023bipartite,lehmann2025consensus}. To use this for large $n$, keys are first mapped to rather large buckets which are then recursively split. Querying such data structures is quite slow as it requires decoding navigation information of variable bit length while descending the resulting tree.

Perhaps the simplest PHF based on \emph{bucket placement} is CHD
\cite{BelazzouguiBD09} which uses a linear bucket assignment function.
However, this immensely slows down construction as, despite processing the buckets in descending size order, the last ones are very difficult to place.
Moreover, variable bit length encoding of seeds is mandatory to achieve constant bits-per-key as the last buckets will need a linear number of trials and thus a logarithmic number of bits for encoding the seed.
FCH~\cite{fox1992faster} and PTHash~\cite{pibiri2021pthash} use a simple nonlinear bucket assignment functions which employs two different expected bucket sizes. This mitigates the slow search and large seed size for the last buckets but does not solve the problem entirely.
PHOBIC \cite{HLPSW24} derives an optimal bucket assignment function which
results in continuously decreasing expected bucket sizes and in expected constant
success probability. However, the nonlinear bucket assignment function slows down
queries. Moreover, this is still not sufficient to achieve linear construction time, and fixed-width encoding of seeds: The former feature requires constant expected bucket size.
The latter would then require $\Omega(\log\log n)$ bits per key as some buckets will require a logarithmic number of trials.
PHast solves this problem by allowing bumping.
CHD, FCH, PTHash and PHOBIC have an additional problem when it comes
to parallel construction. This is achieved by partitioning the input and requires
an additional step in the query to retrieve an offset value.
PHast does not need this as the offset $\lfloor \alpha h(k) \rfloor$ already partitions the input without requiring an additional step.

The closest competitor to PHast is PtrHash \cite{grootkoerkamp:LIPIcs.SEA.2025.21} which was developed independently and with a similar  goal of minimizing query time.
PtrHash can be viewed as a variant of PHOBIC \cite{HLPSW24} engineered for maximum query speed.
PtrHash mitigates the overhead for the nonlinear bucket assignment function by using a simple quadratic or cubic approximation of PHOBIC's optimal choice. It enables fixed-width encoding by replacing the simple greedy search of previous bucket placement approaches by local search: when no feasible seed is found for a bucket, other buckets are ``unplaced'' and pushed back into the list of buckets to be placed. However, for this process to converge, a relatively large amount of nonminimality of the basic construction is needed. The resulting additional query overhead for repairing the MPHF is mitigated by a cache-optimized variant of Elias-Fano coding.
Altogether this results in higher space consumption and slightly slower queries
than PHast. We believe that PtrHash could be improved by adopting the bumping approach of PHast.%
\footnote{Vice-versa PHast can accelerate repairing by adopting cache-optimized Elias-Fano coding from PtrHash. However, this brings no overall query throughput improvement for PHast as repairing is not a bottleneck anyway. Moreover, cache-optimized Elias-Fano introduces space overhead that PHast can better invest elsewhere for accelerating queries.} 


PHast is inspired by other compressed data structures that hash keys to small slices, for example for static function retrieval \cite{Walzer21,DHSW22} or static hash tables~\cite{LSW23slick}. Bumping is adopted from bumbed ribbon retrieval~\cite{DHSW22} and also used in SlickHash~\cite{LSW23slick}.

%% file: method.tex
\section{The PHast Data Structure.}\label{sec:phast}

\input{symbols_tab}

\subsection{Map or Bump Function.}\label{sec:maporbump}

\begin{figure}[tb]
    \centering
    \begin{tikzpicture}[
        frange/.style={matrix of nodes, nodes={draw, inner sep=0, outer sep=0, minimum size=1.4mm}, nodes in empty cells, column sep=-\pgflinewidth},
        buckets/.style={matrix of nodes, nodes={draw, inner sep=0, minimum size=4.2mm, anchor=center}, nodes in empty cells, column sep=-\pgflinewidth},
        f/.style={fill, fill=gray},
        d/.style={fill, fill=darkgray},
        desc/.style={draw, rounded corners, inner sep=2pt, matrix of nodes, column 2/.style={anchor=base west}}
    ]
        
        \matrix[buckets, matrix anchor=west] (B) at (0, 1.5cm) {
            4 & 0 & 3 & 6 &
            2 & 7 & 5 & 1 &
              & 5 &   & 2 &
              &   &   &   & \\
        };
        \node at (B.north west) [above=1ex, anchor=west] {seeds of buckets:};

        \matrix[frange, matrix anchor=west] (FR) at (0, 0) {
            |[f]| & |[f]| & |[f]| & |[f]| & & |[f]| & |[f]| & |[f]| & |[f]| & |[f]| & |[f]| & |[f]| &
            |[f]| & |[f]| & |[f]| & |[f]| & |[f]| & |[f]| & & |[f]| & |[f]| & |[f]| & |[f]| & |[f]| &
            |[f]| & |[f]| & |[f]| & |[f]| &  & |[f]| & |[f]| & & |[f]| & |[f]| & & |[f]|
            & & & & & |[f]| & & & & & &
            & & & & & & & & & \\
        };
        \node at (FR.north west) [above=1ex, anchor=west]  {used/free values:};

        \draw [dashed] (B-1-8.south east) -- (FR-1-24.north east);
        \draw [dashed] (B-1-13.south east) -- (FR-1-44.north east);

        \draw[decorate,decoration={brace,amplitude=2.8ex}] (B-1-8.north east) -- node[above=2.6ex,font=\small]{window of buckets about to be seeded} (B-1-13.north east); 
        \draw[decorate,decoration={brace,mirror,amplitude=1.8ex}] (FR-1-24.south east) -- node[below=1.6ex,font=\small,xshift=-2.2em]{covered by buckets in the window; in the cyclic bitmap} (FR-1-44.south east);

        \matrix[buckets, matrix anchor=west] (B2) at (0, -2.2cm) {
            4 & 0 & 3 & 6 &
            2 & 7 & 5 & 1 &
            \textbf{4} & 5 &   & 2 &
              &   &   &   & \\
        };
        \node at (B2.north west) [above=1ex, anchor=west] {seeds of buckets:};

        \matrix[frange, matrix anchor=west] (FR2) at (0, -3.7cm) {
            |[f]| & |[f]| & |[f]| & |[f]| & & |[f]| & |[f]| & |[f]| & |[f]| & |[f]| & |[f]| & |[f]| &
            |[f]| & |[f]| & |[f]| & |[f]| & |[f]| & |[f]| & & |[f]| & |[f]| & |[f]| & |[f]| & |[f]| &
            |[f]| & |[f]| & |[f]| & |[f]| & |[d]| & |[f]| & |[f]| & |[d]| & |[f]| & |[f]| & &  |[f]|
            & & & & & |[f]| & & & & & &
            & & & & & & & & & \\
        };
        \node at (FR2.north west) [above=1ex, anchor=west]  {used/free values:};

        \draw [dashed] (B2-1-10.south east) -- (FR2-1-30.north east);
        \draw [dashed] (B2-1-15.south east) -- (FR2-1-50.north east);

        \draw[decorate,decoration={brace,amplitude=2.8ex}] (B2-1-10.north east) -- node[above=2.6ex,font=\small]{window of buckets about to be seeded} (B2-1-15.north east);
        \draw[decorate,decoration={brace,mirror,amplitude=1.8ex}] (FR2-1-30.south east) -- node[below=1.6ex,font=\small, xshift=-4.5em]{covered by buckets in the window; in the cyclic bitmap} (FR2-1-50.south east);
        
        \draw[very thick, {Bar}->] ($(B2-1-8.south east)-(0, 0.3cm)$) -- node[left=4mm,font=\tiny,align=right,text width=10em]{shift after seeding the first bucket in the window} ($(B2-1-10.south east)-(0, 0.3cm)$);
    \end{tikzpicture}

\caption[Assigning seeds to buckets]{The seed is assigned to a bucket selected from a window of 5 buckets in the figure (and 256 in our actual implementation). Initially, the window includes buckets with low indexes (on the left in the figure) that cover low function values. Once the seed is assigned to the first (leftmost) bucket in the window, a shift towards higher values (right) is performed, so that the first bucket in the window has no seed (see bottom of figure). To check for collisions, the algorithm uses a cyclic fixed-size bitmap of used/free values from the range covered by the window.}\label{fig:window} 
\end{figure}

PHast is based on the following method, called `\emph{map-or-bump}', which surjectively maps some of the $n$ keys $k_1, \dots, k_n$ to $\{0, 1, \dots, m-1\}$ (we usually use $m=n$), and marks the rest as `\emph{bumped}'.

First, the construction algorithm distributes the keys into $B$ buckets of expected size $\lambda=n/B$. To do this, it maps each key $k_i$ to its $u$-bit \emph{hash code} $c_i = h(k_i)$ and uses integer sorting to group these codes by their \emph{(bucket) index} $\fmap{c_i}{B}$.

Next, it assigns to each bucket, one by one, either a seed from the set $\{1, ..., 2^S-1\}$, or $0$ to indicate bumping. 
Each key $k_i$ in a bucket of seed $s \neq 0$ is mapped to an output value
\begin{equation}\label{eq:eval}
\fmap{c_i}{m-L+1} + p(s, c_i),
\end{equation} 
where:
\begin{itemize}
    \item $L$ is a slice length;    
    (The output set $\{0, \dots, m-1\}$ is covered by $m-L+1$ partially overlapping slices.)
    
    \item $\fmap{c_i}{m-L+1} \in \{0, \dots, m-L\}$ is the first value covered by the slice assigned to $k_i$;
    
    \item $p(s, c_i) \in \{0, \dots, L-1\}$ is the location of $k_i$ in its slice;
    
    \item $p$ is a placement function whose result depends on both of its arguments and is a number from $\{0, \dots, L-1\}$.
    Sections \ref{sec:implementation} and \ref{sec:plus} discuss efficient implementations of $p$.
\end{itemize}
Note that both the bucket index assigned to $k_i$ and the values that can be assigned to $k_i$ (covered by its slice) increase with $c_i$. Table~\ref{tab:symbols} summarizes the notation.

Successive buckets for seed assignment are selected from a window of $W$ (we use $W=256$) consecutive buckets, which is illustrated in Figure~\ref{fig:window}. Initially, the window consists of a nonempty bucket with the lowest index and $W-1$ subsequent buckets. Once the seed is assigned to the first bucket in the window, the window is shifted until its first bucket is unseeded and nonempty.

The unseeded nonempty buckets within the window are kept in a max-priority queue.
The priority increases with the size of the bucket and decreases with its index (see Section \ref{sec:implementation} for details).
This ensures processing buckets in roughly ascending index order, while maintaining a sufficient chance of finding seeds for large buckets.

The selected bucket is assigned a seed which:
\begin{itemize}
    \item is feasible, that is, maps the keys in the bucket to values that do not collide with each other or with values already used by previously assigned buckets; 
    \item minimizes the sum of values it assigns to the keys (and the sum of their $p$ values). Filling low function values is preferable, because they are unavailable to most yet unseeded buckets anyway. 
\end{itemize}

A bitmap of used/free values is employed to check for collisions. Due to the fact that each bucket (and thus the whole window) covers only a certain range of values increasing with the bucket index (shifting the window), the bitmap can contain only the values covered by the window (see Figure \ref{fig:window}).
This makes it efficiently representable with a cyclic buffer of fixed size.
For the same reason, seed assignment can be effectively parallelized, as discussed in Section~\ref{sec:parallel}.

$S$, $L$, $B$, $m$ (and $W$, but we always use $W=256$) are parameters of the method. For independence from $n$, we usually specify $\lambda$ instead of $B$, then $B=\round(n / \lambda)$, while $m$ can be given as a percentage of $n$. 

The \emph{map-or-bump} function is represented as a \emph{seeds} array containing seeds assigned to successive buckets.
Function evaluation for the key $k$ starts with calculating $c=h(k)$ and $s=\seeds[\fmap{c}{B}]$. If $s=0$, $k$ is bumped. Otherwise, its value is calculated by \eqref{eq:eval}.

When $S=8$, \emph{seeds} is an ordinary (thus fast) byte array.
Fast implementations are also possible for $S=4$ and other powers of $2$. However, $S=2$ is too small to be useful, while $S=16$ is too large.

Section \ref{sec:mphf} shows how bumped keys are handled and mapped to free function values, while Sections \ref{sec:implementation} and \ref{sec:plus} discuss implementation details.

\subsection{Minimal and Non-minimal Perfect Hash Functions.}\label{sec:mphf}
A perfect hash function for $n_1$ keys is represented by $r$ \emph{map-or-bump} functions $f_1, \dots, f_r$ such that:
\begin{itemize}
    \item $f_1$ maps $n_1$ input keys to $\{0, \dots, n_1-1\}$,
    \item $f_i$ (for $i=2, \dots, r$) maps $n_{i}$ keys bumped by $f_{i-1}$ to $\{0, \dots, n_i-1\}$,
    \item $f_r$ does not bump any keys.\footnote{To achieve constant worst case query time, we could enforce constant $r$: To avoid bumped keys in layer $r$, choose appropriate parameters (small $\lambda$, large $m$). Note that this would have negligible influence on the overall storage space even if layer $r$ itself is not space efficient.}
\end{itemize}
The construction of a perfect hash function involves building a \emph{map-or-bump} function $f_1$ for all input keys, then $f_2$ for the keys bumped by $f_1$, $f_3$ for those bumped by $f_2$, and so on until no keys are bumped. 
To protect this process from hash code collisions,
$f_1, \dots, f_r$ use different hash functions.

The evaluation of a perfect hash function for key $k$ involves calculating successively $f_1(k)$, $f_2(k)$, $\dots$, $f_i(k)$, where $f_i$ is the first function that does not bump $k$.
If already $f_1$ does not bump $k$ (which is true for $\sim99\%$ of keys for typical configurations) 
then the value of $f_1(k)$ is returned. Otherwise, the returned value is $n_1+\dots+n_{i-1}+f_i(k)$.

To make the perfect function minimal, we use the method proposed in \cite{pibiri2021pthash}.
That is, we use an array $M$ of size $n_2+n_3+\dots+n_r$ that maps successive values greater than or equal to $n_1$ to successive unassigned (due to bumping by $f_1$) values from $\{0, \dots, n_1-1\}$.
The value $n_1+n_2+\dots+n_{i-1}+f_i(k)$ is mapped to $M[n_2+\dots+n_{i-1}+f_i(k)]$. 
With Elias-Fano coding \cite{Elias74,Fano71}, $M$ is stored using space
$\approx \epsilon n(2+\log_2\frac{1}{\epsilon})$, where $\epsilon=\frac{m-n}{m}$.
If $\epsilon$ is small (a~few percent in practice) this implies a~space overhead amounting to a fraction of a~bit per key.

%% file: symbols_tab.tex
\begin{table}
    \caption{Symbols of the main parameters and functions. More details about them can be found in the indicated sections.}\label{tab:symbols}
    \centering
    \begin{tabular}{lll}
        \toprule
            & Meaning & Sections \\
        \midrule
        $S$ & seed size in bits & \ref{sec:maporbump} \\
        $B$ & number of buckets & \ref{sec:maporbump} \\
        $\lambda$ & bucket size; $\lambda=n/B$ & \ref{sec:maporbump}, \ref{sec:benchmark:parameters} \\
        $L$ & slice length & \ref{sec:maporbump}, \ref{sec:benchmark:parameters} \\
        $h$ & $u$-bit hash function; we use $u=64$ & \ref{sec:maporbump}, \ref{sec:implementation} \\
        $p$ & placement function & \ref{sec:maporbump}, \ref{sec:implementation} \\
        $n$ & number of input keys & \ref{sec:maporbump} \\
        $m$ & range of map-or-bump function & \ref{sec:maporbump} \\
        $W$ & window size; we use $W=256$ & \ref{sec:maporbump}, \ref{sec:ablation} \\ 
        \bottomrule
    \end{tabular}
\end{table}


%% file: details.tex
\subsection{Implementation Details.}\label{sec:implementation}
Our implementation can use any $64$-bit hash function ($u=64$).
However, we recommend GxHash \cite{gxhash}. Thanks to SIMD and hardware cryptographic primitives, GxHash is fast and has low collision rate, uniform distribution and high avalanche properties (small changes of input strongly affect its output).

We often take advantage of the fact that for $64$-bit $x$ and $y$, modern CPUs can quickly multiply $x \cdot y$, usually storing in separate registers the more significant (let us denote it by $\hi(x \cdot y)$) and less significant ($\lo(x \cdot y)$) $64$-bit half of the result. Specifically:
\begin{itemize}
    \item $\fmap{x}{r}$ for $u=64$ is (following \cite{fastrange}) calculated as $\hi(x \cdot r)$;
    \item $p(s, c)$ is calculated as \[\hi(\lo(s \cdot 5871781006564002453) \cdot c) \& (L-1),\]
    where: multiplication by $5871781006564002453$ (taken from the FXHash \cite{fxhash} algorithm) spreads the entropy of usually small $s$ across more bits; \emph{bitwise AND} ($\&$) with $L-1$ reduces the result to $\{0, ..., L-1\}$ and is a fast way to compute the remainder of division by $L$, but requires $L$ to be a~power of $2$.
\end{itemize}

\begin{figure}
\begin{tikzpicture}
    \begin{axis}[xlabel=bucket size, ylabel=value, width=\columnwidth, ylabel near ticks, height=.25\textheight, enlarge x limits=0.04, xmax=7.5]
        \addplot[mark=*, dashed] coordinates {
            (1, -50171)
            (2, 59462)
            (3, 109868)
            (4, 141865)
            (5, 163564)
            (6, 181092)
            (7, 192852)
        } node [pos=0.15, yshift=+24pt] {$S=8$};
        \addplot[mark=*, dotted] coordinates {
            (7,192852)
            (8,204612)
            (9,216372)
        };
        \addplot[mark=*, dashed, brown!80!black] coordinates {
            (1, -63000)
            (2, 69496)
            (3, 123197)
            (4, 147274)
            (5, 164471)
            (6, 179677)
            (7, 184910)
        } node [pos=0.79, yshift=+12pt] {$S \geq 11$};
        \addplot[mark=*, dotted, brown!80!black] coordinates {
            (7,184910)
            (8,190143)
        };
        \addplot[mark=*, dashed, blue] coordinates {
            (1,-125171)
            (2,31908)
            (3,74770)
            (4,100065)
            (5,115115)
            (6,126729)
            (7,164878)
        } node [pos=0.8, yshift=-10pt] {$S = 5$};
        \addplot[mark=*, dotted, blue] coordinates {
            (7,164878)
            (8,203027)
        };
    \end{axis}
\end{tikzpicture}

\caption[Size-dependent component of bucket selection priority.]{Plots of $\ell$ functions for selected seed sizes~$S$. The $\ell$ function is the size-dependent component of bucket selection priority. For buckets with more than $7$ keys, it grows linearly, with the same progression as from size $6$ to $7$.}\label{fig:bucketsize}
\end{figure}

We use a binary heap to represent a max-priority queue containing (unseeded and nonempty) buckets within a window from which we select buckets for seed assignment.
Its priority function is $\ell(s) - 1024b$, where $b$ is the bucket index and $\ell$ is an increasing function that modifies the priority based on the size $s$ of the bucket.
The $\ell$ function is represented by an integer array containing the values of $\ell$ for $s$ from $1$ to $7$.
For $s>7$, $\ell$ grows linearly: $\ell(s) = \ell(7) + (s-7) \cdot (\ell(7)-\ell(6))$.

The values in the array representing $\ell$ depend on the size of the seed $S$.
We determined them (separately for different $S$) by numerical optimization (using the Nelder–Mead \cite{NelderMead1965} method), minimizing the sum of bumped keys in a drawn sample of key sets. The values found for different $S$ are similar (Figure \ref{fig:bucketsize} shows examples). However, those for $S<6$ are lower than those for $S \geq 6$. This is because for $S<6$ shorter slices are used (see Section \ref{sec:benchmark:parameters}) 
and thus the bucket index $b$ becomes more important in relation to its size~$s$.
In addition, without significant loss of quality, $\ell(s)$ can be simplified to linear for $s>1$. 
(We have checked that then the size of PHast increases by only about $0.01$ bits/key for $S=8$.)

%% file: plus.tex
\subsection{Faster Construction using \PHastPlus.}\label{sec:plus}

\input{plus_figure}

\PHastPlus{} is a fast to construct variant of PHast that uses one of the following $p$ functions:    
\begin{align}
    p(s, c) &= c \bmod L + s - 1 = c \& (L-1) + s - 1,\label{eq:plus} \\
    p(s, c) &= (c + \delta{}s) \bmod L = (c + \delta{}s) \& (L-1), \label{eq:plus:wrap}
\end{align}
where $\&$ is a \emph{bitwise AND}, $L$ is a power of $2$, $\delta$ is a parameter of the method (we use $\delta \in \{1, 2, 3\}$).

Let $C$ be a multiset of hash codes of keys in the bucket.
When using\footnote{Using \eqref{eq:plus} increases the effective slice length, which should be taken into account by replacing $L$ with $L+2^S-2$ in Equation \eqref{eq:eval} and in the equation calculating the gap's size in Section \ref{sec:parallel}.} \eqref{eq:plus}, the seed $s \in \{1, \dots, 2^S-1\}$ minimizing $\sum_{c \in C}{p(s, c)}$ is quickly found thanks to the following facts:
\begin{itemize}
    \item Since $p$ increases with $s$ ($p(s+1, c) = p(s, c) + 1$ and $\sum_{c \in C}{p(s+1, c)} = \sum_{c \in C}{p(s, c)} + |C|$), it is sufficient to find the smallest feasible seed.
    \item The self-collision test (between keys in the bucket) can be performed once, as its result does not depend on $s$.
    \item The feasibility of multiple ($64$ on modern CPUs) consecutive seeds ($s, s+1, \dots, s+63$) can be checked at once (see Figure \ref{fig:plus}).
    For each $c \in C$, from the bitmap containing $1$ for used (and $0$ for free) values, we read subsequent $64$ bits starting from the value assigned by \eqref{eq:eval} to the pair $(s, c)$.
    The \emph{bitwise OR} of the read words gives a $64$-bit word $r$ containing $0$ on the $i$-th bit if and only if the seed $s+i$ is feasible. The position of the least significant $0$ in $r$ thus indicates the first feasible seed we are looking for. If $r$ does not contain $0$, we continue by checking the next $64$ seeds ($s+64, ..., s+127$) until we reach either feasible or last seed.
\end{itemize}

When using \eqref{eq:plus:wrap}, we search for the optimal seed separately in the (largest possible) intervals $[1, s_1), [s_1, s_2), \dots, [s_I, 2^S)$, such that $\lfloor \frac{c+\delta{}a}{L} \rfloor = \lfloor \frac{c+\delta{}b}{L} \rfloor$ for every $c \in C$ and $a$, $b$ in the same interval.
In each interval, $p$ increases with $s$ ($p(s+1, c)=p(s, c)+\delta{}$ and $\sum_{c \in C}{p(s+1, c)} = \sum_{c \in C}{p(s, c)} + \delta{}|C|$) and, if $\delta=1$,
the same algorithm applies as when using \eqref{eq:plus}.
$\delta{}>1$ are handled by pre-marking illegal shifts with bit ones in $r$, which reduces the parallelism of processing\footnote{Higher parallelism can potentially be recovered by storing free/occupied positions with different remainders from division by $\delta{}$ in separate bitmaps.}.
Typically, raising $\delta{}$ increases the number of intervals and the construction time, but reduces size.

%% file: plus_figure.tex
\begin{figure}
\definecolor{filledColor}{HTML}{999999}
    \begin{tikzpicture}[
        frange/.style={matrix of nodes, nodes={draw, inner sep=0, outer sep=0, minimum size=2.8mm}, nodes in empty cells, column sep=-\pgflinewidth},
        f/.style={fill, fill=filledColor},
        sel/.style={ultra thick, inner sep=0pt, draw, fill, fill opacity=0.15},
    ]
        \matrix[frange] (N) at (0, 0) {
            |[draw opacity=0]| ...\: & |[f]| &
            |[f]| &       & |[f]| & |[f]| &       & |[f]| &       & |[f]| &
            |[f]| & |[f]| &       &
            |[f]| &       & |[f]| & |[f]| &       & |[f]| & |[f]| &       &
                  & |[f]| &       & |[f]| &       & |[f]| &
            |[f]| & |[draw opacity=0]| \:...  \\
        };
        \node at (N.north west) [above=1ex, anchor=west] {bitmap of used/free values:};
        \node[fit={(N-1-3.north west) (N-1-10.south east)}, sel, color=cyan]{};
        \node[fit={(N-1-14.north west) (N-1-21.south east)}, sel, color=magenta, draw opacity=0.7]{};
        \node[fit={(N-1-20.north west) (N-1-27.south east)}, sel, color=green, draw opacity=0.7]{};

        \node (pc0) [below=1.1ex of N-1-3.south, xshift=8mm, font=\footnotesize, color=cyan!50!black] {$\lfloor \alpha c_0 \rfloor + p(s, c_0)$};
        \draw[->, color=cyan!50!black] (pc0.north) + (-8mm, -0.7mm) -> (N-1-3.south);

        \node (pc1) [below=1.1ex of N-1-14.south, xshift=1mm, font=\footnotesize, color=magenta!50!black] {$\lfloor \alpha c_1 \rfloor  + p(s, c_1)$};
        \draw[->, color=magenta!50!black] (pc1.north) + (-1mm, -0.7mm) -> (N-1-14.south);

        \node (pc2) [below=1.1ex of N-1-20.south, xshift=8mm, font=\footnotesize, color=green!50!black] {$\lfloor \alpha c_2 \rfloor + p(s, c_2)$};
        \draw[->, color=green!50!black] (pc2.north) + (-8mm, -0.7mm) -> (N-1-20.south);

        \matrix[frange, nodes={minimum size=4mm}] (N) at (-10mm, -20mm) {
            |[f](Ab)| & & |[f]| & |[f]| &       & |[f]| &       & |[f](Ae)| \\[0.5mm]
            |[f](Bb)| & & |[f]| & |[f]| &       & |[f]| & |[f]| & |(Be)| \\[0.5mm]
            |[f](Cb)| & &       & |[f]| &       & |[f]| &       & |[f](Ce)| \\[1.7mm]
            |[f](R)| & |(F1)| & |[f]| & |[f]| & |(F2)| & |[f]| & |[f]| & |[f]| \\
        };
        \node [left=1mm of Cb, align=center, font=\scriptsize] {bitwise\\OR};
        \node [left=0 of R, align=center, font=\footnotesize] {$r=$};

        \node[fit={(Ab.north west) (Ae.south east)}, sel, color=cyan]{};
        \node[fit={(Bb.north west) (Be.south east)}, sel, color=magenta, draw opacity=0.7]{};
        \node[fit={(Cb.north west) (Ce.south east)}, sel, color=green, draw opacity=0.7]{};

        \draw[thick] ([yshift=-1mm, xshift=-12mm]Cb.south west) -- ([yshift=-1mm]Ce.south east);

        \node[below=3.8ex of F1, red!90!black, font=\footnotesize] (F1f) {$s+1$ is the first feasible seed};
        \draw[->, thick, red!90!black] (F1f) -> (F1);

        \node[below=1.3ex of F2, font=\footnotesize] (F2f) {$s+4$ is feasible};
        \draw[->] (F2f) -> (F2);

        \matrix[matrix of nodes, nodes={font=\footnotesize, inner sep=0, outer sep=0, minimum size=4mm, anchor=center, right}, column sep=1mm, row sep=1mm] at (25mm, -20mm) {
            |[draw, f]| & used (bit 1) \\
            |[draw]| & free (bit 0) \\
        };
    \end{tikzpicture}\caption{\PHastPlus{} tests the feasibility of $8$ ($64$ in our actual implementation) seeds ($s, \dots, s+7$) at once, for a bucket containing hash codes $c_0$, $c_1$, $c_2$.}\label{fig:plus} 
\end{figure}

%% file: parallel.tex
\section{Parallel Construction.}\label{sec:parallel}

We describe a parallelization approach that does not change the query at all. Note that other parallel bucket placement PHFs require an explicit partitioning step that implies overheads at query time.

Hashing keys is trivial to parallelize and sorting hash codes can use well-known techniques for parallel integer sorting.
Thanks to using small slices, bucket placement can also be well paralleled.

\begin{figure}
    \centering
\begin{tikzpicture}[
        frange/.style={matrix of nodes, nodes={draw, inner sep=0, outer sep=0, minimum size=0.88mm}, nodes in empty cells, column sep=-\pgflinewidth},
        buckets/.style={matrix of nodes, nodes={draw, inner sep=0, minimum size=2.7mm, anchor=center}, nodes in empty cells, column sep=-\pgflinewidth},
        desc/.style={draw, rounded corners, inner sep=2pt, matrix of nodes, column 2/.style={anchor=base west}},
        every node/.style={font=\scriptsize}
    ]
        
    \matrix[buckets, matrix anchor=west] (B) at (0, 1.5cm) {
        & & & &
        & & & &
        & & & &
        & & & &
        & & & &
        & & & &
        & & & & \\
    };
    \node at (B.south west) [below=1ex, anchor=west, xshift=1ex] {seeds of buckets};

    \matrix[frange, matrix anchor=west] (FR) at (0, 0) {
        & & & & & & & & & & & &
        & & & & & & & & & & & &
        & & & & & & & & & & & &
        & & & & & & & & & & & &
        & & & & & & & & & & & &
        & & & & & & & & & & & &
        & & & & & & & & & & & &
        & & & & & & & & & \\
        };
    \node at (FR.north west) [above=1ex, xshift=1ex, anchor=west] {used/free values};

    \draw [black!20!blue,dashed] (B-1-1.south west) -- (FR-1-1.north west);
    \draw [black!20!blue,dashed] (B-1-8.south east) -- (FR-1-30.north east);

    \draw [black!50!green,dashed] (B-1-10.south east) -- (FR-1-31.north east);
    \draw [black!50!green,dashed] (B-1-18.south east) -- (FR-1-61.north east);

    \draw [black!50!red,dashed] (B-1-20.south east) -- (FR-1-62.north east);
    \draw [black!50!red,dashed] (B-1-29.south east) -- (FR-1-94.north east);

    \draw[black!20!blue,decorate,decoration={brace,amplitude=1.5ex}] (B-1-1.north west) -- node[above=1.5ex,text width=2.15cm,align=center]{chunk filled by thread 1} (B-1-8.north east);
    \draw[decorate,decoration={brace,amplitude=1.5ex}] (B-1-8.north east) -- node[above=1.5ex]{gap} (B-1-10.north east);
    \draw[black!50!green,decorate,decoration={brace,amplitude=1.5ex},text width=2.15cm,align=center] (B-1-10.north east) -- node[above=1.5ex]{chunk filled by thread 2} (B-1-18.north east);
    \draw[decorate,decoration={brace,amplitude=1.5ex}] (B-1-18.north east) -- node[above=1.5ex]{gap} (B-1-20.north east);
    \draw[black!50!red,decorate,decoration={brace,amplitude=1.5ex}] (B-1-20.north east) -- node[above=1.5ex,text width=2.15cm,align=center]{chunk filled by thread 3} (B-1-29.north east);
\end{tikzpicture}

\caption[Multithreaded seed assigning]{Multithreaded seed assignment. Thanks to the gaps, no communication is required between threads filling separate chunks of the seed array.}\label{fig:mt}
\end{figure}

The seed array is partitioned into $t$ chunks and $t-1$ gaps between the chunks, as shown in Figure~\ref{fig:mt}.
The seeds in chunks are computed in parallel by $t$ threads.
Thanks to the gaps of $\lceil \frac{LB}{m-L+1} \rceil$ buckets each, the function values covered by the chunks do not overlap, and thus no communication between threads is needed.
After the threads have finished computing, the seeds in the gaps are determined.
This can also be done in parallel.
Moreover, the gaps are usually relatively small.
For example, for $n=10^8$ keys, each gap contains about $10^{-5}$ of all buckets.

The only downside of this approach is the lower probability of finding feasible seeds for buckets in the gaps. As long as $n\gg tL$ the resulting additional space overhead is small though. Therefore, when there are very few keys, we reduce the number of threads $t$.

%% file: benchmarks.tex
\section{Benchmarks.}\label{sec:benchmark}
In this section, we check how PHast performs in practice.
All experiments presented herein were conducted using: an AMD Ryzen 5600G @3.9GHz CPU (6 cores, 12 threads, 384KB/3MB/16MB of level 1/2/3 cache), rust 1.84.1 and gcc 12.2.0 compilers (both with third-level and CPU native optimizations enabled), 12 threads in the case of multithreaded computing.

Our PHast implementation is developed in Rust as part of the open-source \emph{BSuccinct} \cite{FigShare,beling2024BSuccinct,BSuccinctGit} package.
We use GxHash \cite{gxhash} as hash function.
Unless otherwise specified, we encode the array making PHast minimal (denoted as $M$ in Section \ref{sec:mphf}) using the Elias-Fano \cite{Elias74,Fano71} method. 

\subsection{Parameter Tuning.}\label{sec:benchmark:parameters}

\input{plot_parameter_phast}

\input{plot_parameter_plus}

\input{plot_parameter_pluswrap}

\input{plot_parameter_pluswrap2}

\input{plot_parameter_pluswrap3}

We selected values for many PHast parameters empirically.
Here we present the conclusions and main results of our experiments, including the optimal (minimizing the size) parameter values.

We found that $L$ (slice length) can, without significant performance degradation, be a power of $2$. This allowed us to optimize the implementation of the $p$ function (see Sections~\ref{sec:implementation} and \ref{sec:plus}).
However, the optimal $L$ depends on $S$ (seed size in bits).

For large $n$ (number of keys), regular PHast uses $L=2048$ for $S \geq 12$, $L=1024$ for $S \in \{6, \dots, 11\}$ and $L=512$ for $S < 6$.
As $n$ decreases, we reduce $L$: to $512$ for $n < 140000$, $256$ for $n<12000$, $128$ for $n<9500$, $64$ for $n<1300$, a power of $2$ not exceeding $n$ for $n<64$.
$L$ can never be greater than $m$ (the number of function values), and we use $m=n$.

\begin{table}
  \setlength{\tabcolsep}{5pt}
  \caption{Values of $L$ used with \PHastPlus{} \eqref{eq:plus:wrap} $\delta \in \{1, 2, 3\}$, $S \in \{8, \dots, 12\}$ (for $\delta=2$, we use $L \leq 11$).}\label{tab:PHastPlus:wrap:L}
  \begin{tabular}{cccccc}
  \toprule
  $\delta$ & 8 & 9 & 10 & 11 & 12 \\
  \midrule
  1 & 512, 1024 & 1024 & 1024, 2048 & 2048 & 4096 \\
  2 & 1024 & 1024, 2048 & 2048 & 4096 & n/a \\
  3 & 1024 & 2048 & 2048, 4096 & 4096 & 4096 \\
  \bottomrule
  \end{tabular}
\end{table}

We usually use \PHastPlus{} with $S$ from $8$ (smaller ones do not measurably speed up construction) to $12$ (larger ones require very large $L$ and give only a negligible reduction in size).
We use \PHastPlus{} \eqref{eq:plus} with $L=2^{S+1}$ and \PHastPlus{} \eqref{eq:plus:wrap} with $\delta \in \{1, 2, 3\}$ and the values of $L$ given in Table \ref{tab:PHastPlus:wrap:L}.

When the number of keys $n$ drops below $8192$ (for \eqref{eq:plus}) or $4096$ (for \eqref{eq:plus:wrap}), we build the final layer without bumping, using regular PHast $S=8$, $\lambda=4$, $m=1.2n$.


The optimal $\lambda$ (expected bucket size) also depends on $S$. We examined this dependence for various $S$ and all $\lambda$ from $2$ to $8$ in $0.05$ increments, using $10^8$ random $64$-bit integer keys.
Figures \ref{fig:optimal:lambda}, \ref{fig:optimal:lambda:plus}, \ref{fig:optimal:lambda:plus1}, \ref{fig:optimal:lambda:plus2}, \ref{fig:optimal:lambda:plus3} show the results, including $\lambda$ values that minimize the size.
(Note that $\lambda$ values slightly lower than those listed there can noticeably speed up queries.)
A natural conjecture is that the space consumption of PHast -- as that of many other bucket placement approaches -- is $\log_2 e+\mathrm{O}(\log(\lambda)/\lambda)$ asymptotically \cite{HLPSW24}.
We therefore fitted such a function to the optimal $\lambda$ and shown it in the figures.


\subsection{Comparison to Other Methods.}\label{sec:benchmark:comparision}

\begin{figure*}
\input{ST50M}

\caption[Trade-offs between space and speed of construction and querying for $5 \cdot 10^7$ keys.]{Trade-offs between space and speed of querying and (1-threaded) construction for $5 \cdot 10^7$ keys. Each top plot shows the Pareto frontiers of two properties (ignoring the third and configurations inferior in one property and no better in another than any different configuration of the same method).
Each bottom plot shows the relationships between all three properties. For each combination of two of them, it indicates by color the algorithm offering the best value for the third.}\label{plot:ST50M}
\end{figure*}

\begin{figure*}
\input{MT50M}

\caption[Trade-offs between space and speed of construction and querying for $5 \cdot 10^7$ keys.]{Trade-offs between space and speed of querying and (12-threaded) construction for $5 \cdot 10^7$ keys. (Note that FiPS, unlike the other methods shown, does not support and thus does not use multiple threads for construction.) Each top plot shows the Pareto frontiers of two properties (ignoring the third and configurations inferior in one property and no better in another than any different configuration of the same method).
Each bottom plot shows the relationships between all three properties. For each combination of two of them, it indicates by color the algorithm offering the best value for the third.}\label{plot:MT50M}
\end{figure*}



\begin{table*}
    \centering
    \caption{Performance of various methods on $5 \cdot 10^7$ keys.}\label{tab:50M:extra}
    \input{table-50M-extra}    
\end{table*}

\begin{table*}
    \centering
    \caption{Performance of various methods on $5 \cdot 10^8$ keys.}\label{tab:500M:extra}
    \input{table-500M-extra}    
\end{table*}




Figures \ref{plot:ST50M}, \ref{plot:MT50M} and Tables \ref{tab:50M:extra}, \ref{tab:500M:extra} compare the performance of the following methods reviewed in Section \ref{sec:related}, based on:
\begin{itemize}
    \item bucket placement: PHast and \PHastPlus{} (this paper), PtrHash \cite{grootkoerkamp:LIPIcs.SEA.2025.21}, PTHash \cite{pibiri2021pthash}, PHOBIC \cite{HLPSW24}; 
    \item recursive splitting: SIMDRecSplit \cite{bez2023high}, 
    Bipartite ShockHash and Bipartite ShockHash-Flat \cite{lehmann2023bipartite};
    \item fingerprinting: FiPS \cite{Lehmann24a}, FMPH and FMPHGO~\cite{beling2023fingerprinting};
    \item static function retrieval: SicHash \cite{dillinger2022burr}.
\end{itemize}
Benchmarked implementations and designations of their parameters come from the cited papers and their authors. 
For the sake of readability, we excluded from our tests some methods that perform worse than their improved counterparts, notably RecSplit \cite{esposito2020recsplit} (due to SIMDRecSplit), ShockHash \cite{lehmann2023shockhash} (due to Bipartite ShockHash), and BBHash \cite{limasset2017fast} (due to FMPH).

Exploiting the ease of choosing a hash algorithm for methods implemented in Rust, we use the same GxHash \cite{gxhash} for all PHast variants, PtrHash, FMPH and FMPHGO. 
We test PHast with two encodings of the array ensuring it is minimal (see Section \ref{sec:mphf}): Elias-Fano (EF) and Compact (C; which stores each value using the same, lowest possible number of bits).

To perform the benchmarks, we adopted the \emph{MPHF-Experiments} measurement framework~\cite{Lehmann:MPHFExperiments,FigShare} developed by Hans-Peter Lehmann and used in many recent papers. This implies the use of random string keys, each of uniformly random length between $10$ and $50$ bytes.

For readability, since the impact of multithreaded construction on query speed and size is small,
these parameters in the tables are averages 
of the values obtained for single- and multithreaded construction.

\paragraph{Conclusions:} Algorithms based on bucket placement do not offer as small (close to the theoretical minimum) size as those based on recursive splitting, but instead they feature much shorter evaluation and, often, construction times.

Among them, the PHast family and PtrHash feature particularly fast evaluation, well ahead of all other methods.
Of the two, the PHast family offers lower space consumption and faster evaluation for a small ($5 \cdot 10^7$) set of keys, for which its advantage comes from performing fewer operations.
For $5 \cdot 10^8$ keys, the evaluation speed becomes dominated by memory accesses, and the difference blurs. PtrHash catches up and seems to overtake \PHastPlus{} \eqref{eq:plus} (which usually bumps the most keys and thus performs the most memory reads per evaluation in the PHast family). 

\PHastPlus{} construction is very fast, competitive with PtrHash, SIMDRecSplit and FiPS, slower only than the (much larger) FMPH and only for a smaller set of keys. Moreover, \PHastPlus{} usually offers the best space/construction speed trade-off, which, combined with its very fast evaluation, makes it very versatile and dominant over many other methods in practice.


Regular PHast does not offer such fast construction, but it can make very good (relatively better than other methods) use of multithreading. This confirms that seed assignment, which contributes much more to the construction time of PHast than \PHastPlus{}, is efficiently parallelized (by the method described in Section \ref{sec:parallel}). 

As can be inferred by comparing the results for $5 \cdot 10^7$ and $5 \cdot 10^8$ keys, the PHast family belongs to methods whose construction time scales almost linearly with the size of the input data (as a result, \PHastPlus{} outperforms FMPH for $5 \cdot 10^8$ keys).
This is the effect of 
CPU-cache friendliness (filling the seed roughly sequentially and using cyclic buffers results in close addresses of subsequent memory accesses).


Due to recent progress of bucket placement methods, those based on fingerprints and SicHash appear to be losing their appeal.
Note that recent improvements to the FMPH and FMPHGO implementations and their use with GxHash have made them perform relatively a bit better than in older papers.

\input{construction_steps}

\input{ablation_study}

%% file: plot_parameter_phast.tex
\begin{figure*}
\begin{center}
\begin{tabular}{lccccccccc}
  \toprule
  seed size $S$ [bit] & 4 & 5 & 6 & 7 & 8 & 9 & 10 & 11 & 12 \\
  \midrule
  bucket size $\lambda$ minimizing space & 2.9 & 3.2 & 3.6 & 4.05 & 4.7 & 5.4 & 6.05 & 6.75 & 7.35 \\
  minimum size [bit/key] & 2.32 & 2.16 & 2.04 & 1.96 & 1.91 & 1.88 & 1.85 & 1.83 & 1.81 \\
  \bottomrule
\end{tabular}
\end{center}

\pgfplotstableread{plots_data/parameter_phast/phast_4.csv}{\sizeA}
\pgfplotstableread{plots_data/parameter_phast/phast_5.csv}{\sizeB}
\pgfplotstableread{plots_data/parameter_phast/phast_6.csv}{\sizeC}
\pgfplotstableread{plots_data/parameter_phast/phast_7.csv}{\sizeD}
\pgfplotstableread{plots_data/parameter_phast/phast_8.csv}{\sizeE}
\pgfplotstableread{plots_data/parameter_phast/phast_9.csv}{\sizeF}
\pgfplotstableread{plots_data/parameter_phast/phast_10.csv}{\sizeG}
\pgfplotstableread{plots_data/parameter_phast/phast_11.csv}{\sizeH}
\pgfplotstableread{plots_data/parameter_phast/phast_12.csv}{\sizeI}
\pgfplotstableread{plots_data/parameter_phast/phast_min.csv}{\approx}

\begin{tikzpicture}
\begin{axis}[xlabel={$\lambda$}, axis lines=left, enlarge y limits=auto, xmin=2, xmax=8, width=0.8\textwidth, height=0.4\textheight, scale only axis, every axis x label/.style={at={(current axis.right of origin)},anchor=west}, ylabel={Space [bit/key]}, ymin=1.78, ymax=2.6, domain=2:8,samples=120, xtick distance=1, minor tick num=3]
	\addplot[color=blue] table [x expr={\thisrow{bucket_size100} * 0.01}, y={bits/key}] {\sizeA} node [pos=0.53, yshift=-16pt] {$S=4$};
	\addplot[color=blue!75!red] table [x expr={\thisrow{bucket_size100} * 0.01}, y={bits/key}] {\sizeB} node [pos=0.1, yshift=-20pt] {$S=5$};
    \addplot[color=blue!50!red] table [x expr={\thisrow{bucket_size100} * 0.01}, y={bits/key}] {\sizeC} node [pos=0.8, yshift=16pt] {$S=6$};
    \addplot[color=blue!25!red] table [x expr={\thisrow{bucket_size100} * 0.01}, y={bits/key}] {\sizeD} node [pos=0.8, yshift=16pt] {$S=7$};
    \addplot[color=red] table [x expr={\thisrow{bucket_size100} * 0.01}, y={bits/key}] {\sizeE} node [pos=0.84, yshift=14pt] {$S=8$};
    \addplot[color=red!75!green] table [x expr={\thisrow{bucket_size100} * 0.01}, y={bits/key}] {\sizeF} node [pos=0.88, yshift=12pt] {$S=9$};
    \addplot[color=red!50!green] table [x expr={\thisrow{bucket_size100} * 0.01}, y={bits/key}] {\sizeG} node [pos=0.94, yshift=11pt] {$S=10$};
    \addplot[color=red!25!green] table [x expr={\thisrow{bucket_size100} * 0.01}, y={bits/key}] {\sizeH} node [pos=0.97, yshift=8pt] {$S=11$};
    \addplot[color=green] table [x expr={\thisrow{bucket_size100} * 0.01}, y={bits/key}] {\sizeI} node [pos=0.63, xshift=20pt] {$S=12$};
    \addplot[only marks,mark=*,mark options={fill=black},color=black] table [
        x expr={\thisrow{bucket_size100} * 0.01}, y={bits/key}
    ] {\approx};
    \addplot[color=gray, dashed]  expression {
        ln(2)^(-1) + 0.955 * (ln(2)^(-1)) * ln(x)/x
    } node [pos=0.15, yshift=-10pt] {$\log_2e + 0.955\log_2(\lambda)/\lambda$};
\end{axis}
\end{tikzpicture}
\begin{tikzpicture}
\pgfplotsset{
        y axis style/.style={
            yticklabel style=#1,
            ylabel style=#1,
            y axis line style=#1,
            ytick style=#1
       }
}

\begin{axis}[
    axis y line*=left,
    ylabel={Construction time [ns/key]}, xlabel={$\lambda$},
    width=0.8\textwidth, height=0.25\textheight, scale only axis,
    xmin=2, xmax=8,
    y axis style=blue!75!black,
    scale only axis, every axis x label/.style={at={(current axis.right of origin)},anchor=west},
    xtick distance=1, minor tick num=3,
    extra x ticks={4.7},
    extra x tick style={
        grid=major,major grid style={gray,dashed},
        tick label style={text=gray, yshift=0.22\textheight, xshift=-1ex, rotate=90}
    },
    extra x tick labels={$S=8$ minimum size}
  ]
  \addplot[blue] table [
    x expr={\thisrow{bucket_size100} * 0.01},
    y={build_ns/key}, 
    col sep=space
  ] {\sizeE};
\end{axis}

\begin{axis}[
    axis y line*=right,
    ylabel near ticks,
    width=0.8\textwidth, height=0.25\textheight, scale only axis,
    axis x line=none,
    ylabel={Query time [ns/query]},
    xmin=2, xmax=8,
    y axis style=red!75!black
]
\addplot[
    red,
] table [
    x expr={\thisrow{bucket_size100} * 0.01},
    y={query_ns/key},
    col sep=space
] {\sizeE};
\end{axis}
\end{tikzpicture}

\caption[Optimal PHast bucket size selection for different seed sizes.]{
The top plot shows how PHast size depends on bucket size ($\lambda$), for seed sizes ($S$) from $4$ to $12$. The black dots indicate the minimum sizes, which are also given in the table. The bottom plot illustrates the impact of $\lambda$ on construction and query times for $S=8$. To speed up queries, one can use $\lambda$ somewhat lower than this minimizing size.}\label{fig:optimal:lambda}
\end{figure*}

%% file: plot_parameter_plus.tex
\begin{figure}
\begin{tabular}{lccccc}
  \toprule
  seed size $S$ [bit] & 8 & 9 & 10 & 11 & 12 \\
  \midrule
  $\lambda$ minimizing size & 5.25 & 5.55 & 6.10 & 6.70 & 7.05 \\
  size [bit/key] & 2.12 & 2.03 & 1.98 & 1.95 & 1.94 \\
  \bottomrule
\end{tabular}
\vspace{0.25cm}
\pgfplotstableread{plots_data/parameter_plus/plus_8.csv}{\sizeE}
\pgfplotstableread{plots_data/parameter_plus/plus_9.csv}{\sizeF}
\pgfplotstableread{plots_data/parameter_plus/plus_10.csv}{\sizeG}
\pgfplotstableread{plots_data/parameter_plus/plus_11.csv}{\sizeH}
\pgfplotstableread{plots_data/parameter_plus/plus_12.csv}{\sizeI}
\pgfplotstableread{plots_data/parameter_plus/plus_min.csv}{\approx}

\begin{tikzpicture}
\begin{axis}[xlabel={$\lambda$}, axis lines=left, enlarge y limits=auto, xmin=4.5, xmax=8, width=0.8\columnwidth, height=0.25\textheight, scale only axis, every axis x label/.style={at={(current axis.right of origin)},anchor=west}, ylabel={Space [bit/key]}, ymin=1.88, ymax=2.4, domain=2:8,samples=120, xtick distance=1, minor tick num=3]
    \addplot[color=red] table [x expr={\thisrow{bucket_size100} * 0.01}, y={bits/key}] {\sizeE} node [pos=0.8, yshift=12pt] {$S=8$};
    \addplot[color=red!75!green] table [x expr={\thisrow{bucket_size100} * 0.01}, y={bits/key}] {\sizeF} node [pos=0.8, yshift=12pt] {$S=9$};
    \addplot[color=red!50!green] table [x expr={\thisrow{bucket_size100} * 0.01}, y={bits/key}] {\sizeG} node [pos=0.9, yshift=12pt] {$S=10$};
    \addplot[color=red!25!green] table [x expr={\thisrow{bucket_size100} * 0.01}, y={bits/key}] {\sizeH} node [pos=0.95, yshift=10pt] {$S=11$};
    \addplot[color=green] table [x expr={\thisrow{bucket_size100} * 0.01}, y={bits/key}] {\sizeI} node [pos=0.65, xshift=20pt] {$S=12$};
    \addplot[only marks,mark=*,mark options={fill=black},color=black] table [
        x expr={\thisrow{bucket_size100} * 0.01}, y={bits/key}
    ] {\approx};
    \addplot[color=gray, dashed]  expression {
        ln(2)^(-1) + 1.247 * (ln(2)^(-1)) * ln(x)/x
    } node [pos=0.6, yshift=-22pt] {$\log_2e + 1.247\log_2(\lambda)/\lambda$};
\end{axis}
\end{tikzpicture}
	



\caption[Optimal \PHastPlus{} \eqref{eq:plus} bucket size selection for different seed sizes.]{Plot of \PHastPlus{} \eqref{eq:plus} size dependence on bucket size ($\lambda$), for $S$ from $8$ to $12$. The black dots indicate the minimum sizes, which are also given in the table.}\label{fig:optimal:lambda:plus}
\end{figure}

%% file: plot_parameter_pluswrap.tex
\begin{figure}
\begin{tabular}{lccccc}
  \toprule
  seed size $S$ [bit] & 8 & 9 & 10 & 11 & 12 \\
  \midrule
  $\lambda$ minimizing size & 5.35 & 5.75 & 6.35 & 6.85 & 7.4 \\
  size [bit/key] & 2.09 & 1.96 & 1.90 & 1.84 & 1.82 \\
  \bottomrule
\end{tabular}
\vspace{0.25cm}
\pgfplotstableread{plots_data/parameter_pluswrap/1pluswrap2_8.csv}{\sizeE}
\pgfplotstableread{plots_data/parameter_pluswrap/1pluswrap2_9.csv}{\sizeF}
\pgfplotstableread{plots_data/parameter_pluswrap/1pluswrap2_10.csv}{\sizeG}
\pgfplotstableread{plots_data/parameter_pluswrap/1pluswrap2_11.csv}{\sizeH}
\pgfplotstableread{plots_data/parameter_pluswrap/1pluswrap2_12.csv}{\sizeI}
\pgfplotstableread{plots_data/parameter_pluswrap/1pluswrap2_min.csv}{\approx}

\begin{tikzpicture}
\begin{axis}[xlabel={$\lambda$}, axis lines=left, enlarge y limits=auto, xmin=4.5, xmax=8, width=0.8\columnwidth, height=0.25\textheight, scale only axis, every axis x label/.style={at={(current axis.right of origin)},anchor=west}, ylabel={Space [bit/key]}, ymin=1.75, ymax=2.4, domain=2:8,samples=120, xtick distance=1, minor tick num=3]
    \addplot[color=red] table [x expr={\thisrow{bucket_size100} * 0.01}, y={bits/key}] {\sizeE} node [pos=0.8, yshift=12pt] {$S=8$};
    \addplot[color=red!75!green] table [x expr={\thisrow{bucket_size100} * 0.01}, y={bits/key}] {\sizeF} node [pos=0.8, yshift=10pt] {$S=9$};
    \addplot[color=red!50!green] table [x expr={\thisrow{bucket_size100} * 0.01}, y={bits/key}] {\sizeG} node [pos=0.9, yshift=10pt] {$S=10$};
    \addplot[color=red!25!green] table [x expr={\thisrow{bucket_size100} * 0.01}, y={bits/key}] {\sizeH} node [pos=0.95, yshift=10pt] {$S=11$};
    \addplot[color=green] table [x expr={\thisrow{bucket_size100} * 0.01}, y={bits/key}] {\sizeI} node [pos=0.65, xshift=20pt] {$S=12$};
    \addplot[only marks,mark=*,mark options={fill=black},color=black] table [
        x expr={\thisrow{bucket_size100} * 0.01}, y={bits/key}
    ] {\approx};
    \addplot[color=gray, dashed]  expression {
        ln(2)^(-1) + 0.983 * (ln(2)^(-1)) * ln(x)/x
    } node [pos=0.6, yshift=-18pt] {$\log_2e + 0.983\log_2(\lambda)/\lambda$};
\end{axis}
\end{tikzpicture}
	



\caption[Optimal \PHastPlus{} \eqref{eq:plus:wrap} $\delta=1$ bucket size selection for different seed sizes.]{
Plot of \PHastPlus{} \eqref{eq:plus:wrap} $\delta=1$ size dependence on bucket size ($\lambda$), for $S=$ $8$ (with $L=1024$), $9$, $10$ (with $L=2048$), $11$, $12$. The black dots indicate the minimum sizes, which are also given in the table.}\label{fig:optimal:lambda:plus1}
\end{figure}

%% file: plot_parameter_pluswrap2.tex
\begin{figure}
\begin{tabular}{lcccc}
  \toprule
  seed size $S$ [bit] & 8 & 9 & 10 & 11 \\
  \midrule
  $\lambda$ minimizing size & 5.05 & 5.65 & 6.15 & 6.8 \\
  minimum size [bit/key] & 2.01 & 1.94 & 1.87 & 1.85 \\
  \bottomrule
\end{tabular}
\vspace{0.25cm}
\pgfplotstableread{plots_data/parameter_pluswrap/2pluswrap2_8.csv}{\sizeE}
\pgfplotstableread{plots_data/parameter_pluswrap/2pluswrap2_9.csv}{\sizeF}
\pgfplotstableread{plots_data/parameter_pluswrap/2pluswrap2_10.csv}{\sizeG}
\pgfplotstableread{plots_data/parameter_pluswrap/2pluswrap2_11.csv}{\sizeH}
\pgfplotstableread{plots_data/parameter_pluswrap/2pluswrap2_12.csv}{\sizeI}
\pgfplotstableread{plots_data/parameter_pluswrap/2pluswrap2_min.csv}{\approx}

\begin{tikzpicture}
\begin{axis}[xlabel={$\lambda$}, axis lines=left, enlarge y limits=auto, xmin=3, xmax=8, width=0.7\columnwidth, height=0.22\textheight, scale only axis, every axis x label/.style={at={(current axis.right of origin)},anchor=west}, ylabel={Space [bit/key]}, ymin=1.8, ymax=2.3, domain=2:8,samples=120, xtick distance=1, minor tick num=3]
    \addplot[color=red] table [x expr={\thisrow{bucket_size100} * 0.01}, y={bits/key}] {\sizeE} node [pos=0.8, yshift=16pt] {$S=8$};
    \addplot[color=red!66!green] table [x expr={\thisrow{bucket_size100} * 0.01}, y={bits/key}] {\sizeF} node [pos=0.8, yshift=14pt] {$S=9$};
    \addplot[color=red!33!green] table [x expr={\thisrow{bucket_size100} * 0.01}, y={bits/key}] {\sizeG} node [pos=0.9, yshift=16pt] {$S=10$};
    \addplot[color=green] table [x expr={\thisrow{bucket_size100} * 0.01}, y={bits/key}] {\sizeH} node [pos=0.6, xshift=20pt] {$S=11$};
    \addplot[only marks,mark=*,mark options={fill=black},color=black] table [
        x expr={\thisrow{bucket_size100} * 0.01}, y={bits/key}
    ] {\approx};
    \addplot[color=gray, dashed]  expression {
        ln(2)^(-1) + (ln(2)^(-1)) * ln(x)/x
    } node [pos=0.4, yshift=-18pt] {$\log_2e + \log_2(\lambda)/\lambda$};
\end{axis}
\end{tikzpicture}
	
\begin{tikzpicture}
\pgfplotsset{
        y axis style/.style={
            yticklabel style=#1,
            ylabel style=#1,
            y axis line style=#1,
            ytick style=#1
       }
}

\begin{axis}[
    axis y line*=left,
    ylabel={Construction time [ns/key]}, xlabel={$\lambda$},
    width=0.7\columnwidth, height=0.22\textheight, scale only axis,
    xmin=3, xmax=8,
    y axis style=blue!75!black,
    scale only axis, every axis x label/.style={at={(current axis.right of origin)},anchor=west},
    xtick distance=1, minor tick num=3,
    extra x ticks={5.05},
    extra x tick style={
        grid=major,major grid style={gray,dashed},
        tick label style={text=gray, yshift=0.21\textheight, xshift=-1ex, rotate=90}
    },
    extra x tick labels={$S=8$ minimum size}
  ]
  \addplot[blue] table [
    x expr={\thisrow{bucket_size100} * 0.01},
    y={build_ns/key}, 
    col sep=space
  ] {\sizeE};
\end{axis}

\begin{axis}[
    axis y line*=right,
    ylabel near ticks,
    width=0.7\columnwidth, height=0.22\textheight, scale only axis,
    axis x line=none,
    ylabel={Query time [ns/query]},
    xmin=3, xmax=8,
    y axis style=red!75!black
]
\addplot[
    red,
] table [
    x expr={\thisrow{bucket_size100} * 0.01},
    y={query_ns/key},
    col sep=space
] {\sizeE};
\end{axis}

\end{tikzpicture}

\caption[Optimal \PHastPlus{} \eqref{eq:plus:wrap} $\delta=2$ bucket size selection for different seed sizes.]{
The top plot shows how \PHastPlus{} \eqref{eq:plus:wrap} $\delta=2$ size depends on bucket size ($\lambda$), for $S=$ $8$, $9$ (with $L=2048$), $10$, $11$. The black dots indicate the minimum sizes, which are also given in the table. The bottom plot illustrates the impact of $\lambda$ on construction and query times for $S=8$. To speed up queries, one can use $\lambda$ somewhat lower than this minimizing size.}\label{fig:optimal:lambda:plus2}
\end{figure}

%% file: plot_parameter_pluswrap3.tex
\begin{figure}
\begin{tabular}{lccccc}
  \toprule
  seed size $S$ [bit] & 8 & 9 & 10 & 11 & 12 \\
  \midrule
  $\lambda$ minimizing size & 5.00 & 5.65 & 6.15 & 6.80 & 7.45 \\
  size [bit/key] & 1.97 & 1.91 & 1.87 & 1.85 & 1.82 \\
  \bottomrule
\end{tabular}
\vspace{0.25cm}
\pgfplotstableread{plots_data/parameter_pluswrap/3pluswrap2_8.csv}{\sizeE}
\pgfplotstableread{plots_data/parameter_pluswrap/3pluswrap2_9.csv}{\sizeF}
\pgfplotstableread{plots_data/parameter_pluswrap/3pluswrap2_10.csv}{\sizeG}
\pgfplotstableread{plots_data/parameter_pluswrap/3pluswrap2_11.csv}{\sizeH}
\pgfplotstableread{plots_data/parameter_pluswrap/3pluswrap2_12.csv}{\sizeI}
\pgfplotstableread{plots_data/parameter_pluswrap/3pluswrap2_min.csv}{\approx}

\begin{tikzpicture}
\begin{axis}[xlabel={$\lambda$}, axis lines=left, enlarge y limits=auto, xmin=4.5, xmax=8, width=0.8\columnwidth, height=0.25\textheight, scale only axis, every axis x label/.style={at={(current axis.right of origin)},anchor=west}, ylabel={Space [bit/key]}, ymin=1.8, ymax=2.3, domain=2:8,samples=120, xtick distance=1, minor tick num=3]
    \addplot[color=red] table [x expr={\thisrow{bucket_size100} * 0.01}, y={bits/key}] {\sizeE} node [pos=0.8, yshift=14pt] {$S=8$};
    \addplot[color=red!75!green] table [x expr={\thisrow{bucket_size100} * 0.01}, y={bits/key}] {\sizeF} node [pos=0.9, yshift=12pt] {$S=9$};
    \addplot[color=red!50!green] table [x expr={\thisrow{bucket_size100} * 0.01}, y={bits/key}] {\sizeG} node [pos=0.92, yshift=14pt] {$S=10$};
    \addplot[color=red!25!green] table [x expr={\thisrow{bucket_size100} * 0.01}, y={bits/key}] {\sizeH} node [pos=0.95, yshift=10pt] {$S=11$};
    \addplot[color=green] table [x expr={\thisrow{bucket_size100} * 0.01}, y={bits/key}] {\sizeI} node [pos=0.67, xshift=20pt] {$S=12$};
    \addplot[only marks,mark=*,mark options={fill=black},color=black] table [
        x expr={\thisrow{bucket_size100} * 0.01}, y={bits/key}
    ] {\approx};
    \addplot[color=gray, dashed]  expression {
        ln(2)^(-1) + 0.989*(ln(2)^(-1)) * ln(x)/x
    } node [pos=0.6, yshift=-16pt] {$\log_2e + 0.989\log_2(\lambda)/\lambda$};
\end{axis}
\end{tikzpicture}
	




\caption[Optimal \PHastPlus{} \eqref{eq:plus:wrap} $\delta=3$ bucket size selection for different seed sizes.]{
Plot of \PHastPlus{} \eqref{eq:plus:wrap} $\delta=3$ size dependence on bucket size ($\lambda$), for $S=$ $8$, $9$, $10$ ($L=2048$), $11$, $12$. The black dots indicate the minimum sizes, which are also given in the table.}\label{fig:optimal:lambda:plus3}
\end{figure}

%% file: table-50M-extra.tex
\newcommand{\mr}[1]{\multirow{2}{*}{#1}}
\begin{centering}\small
\begin{tabular}[t]{ll rrrrr}
    \toprule
    Method & Configuration & Space      & Query      & \multicolumn{3}{c}{Construction [ns/key]} \\
             &               & {\scriptsize [bit/key]} & {\scriptsize [ns/query]} & {\scriptsize 1 thread} & {\scriptsize 12 threads} & {\scriptsize speedup} \\ \midrule
                \mr{PHOBIC}& $\lambda$=$3.9$, $\alpha$=$1.0$, IC-C & 3.17 &  42 &  232 &  33 & 7.0 \\
                           & $\lambda$=$6.5$, $\alpha$=$1.0$, IC-C & 2.33 &  33 &  953 & 115 & 8.3 \\
                           & $\lambda$=$4.5$, $\alpha$=$1.0$, IC-R & 2.12 &  66 &  288 &  39 & 7.3 \\
                           & $\lambda$=$6.5$, $\alpha$=$1.0$, IC-R & 1.86 &  60 &  954 & 115 & 8.3 \\
                           & $\lambda$=$9.0$, $\alpha$=$1.0$, IC-R & 1.75 &  57 & 7\,551 & 889 & 8.5 \\
                           \addlinespace[0.3em]
                \mr{PTHash}& $\lambda$=$4.0$, $\alpha$=$0.99$, C-C & 3.19 &  40 &  322 & 152 & 2.1 \\
                            & $\lambda$=$5.0$, $\alpha$=$0.99$, EF & 2.11 &  54 &  577 & 231 & 2.5 \\
                            \addlinespace[0.3em]
            \mr{PTHash-HEM}& $\lambda$=$4.0$, $\alpha$=$0.99$, C-C & 3.08 &  44 &  333 &  42 & 8.0 \\
                            & $\lambda$=$5.0$, $\alpha$=$0.99$, EF & 2.11 &  59 &  580 &  67 & 8.6 \\
                            \addlinespace[0.3em]
                    \mr{FMPH}& $\gamma$=$2.0$ & 3.40 &  69 &   40 &  10 & 4.1 \\
                            & $\gamma$=$1.0$ & 2.80 &  76 &   57 &  16 & 3.5 \\
                            \addlinespace[0.3em]
                        \mr{FMPHGO}& $\gamma$=$2.0, s$=$4, b$=$16$ & 2.86 &  45 &  329 &  48 & 6.8 \\
                                   & $\gamma$=$1.0, s$=$4, b$=$16$ & 2.21 &  52 &  158 &  48 & 3.3 \\
                                   \addlinespace[0.3em]
                          \mr{PHast}& $S$=$4$, $\lambda$=$2.6$, EF & 2.33 &  22 &  133 &  25 & 5.2 \\
                                    & $S$=$5$, $\lambda$=$2.8$, EF & 2.18 &  15 &  167 &  27 & 6.2 \\
                          & $S$=$6$, $\lambda$=$3.2$, EF & 2.08 &  12 &  241 &  33 & 7.2 \\
                                     & $S$=$8$, $\lambda$=$4.2$, C & 2.05 &   9 &  624 &  71 & 8.8 \\
                                    & $S$=$7$, $\lambda$=$3.7$, EF & 2.00 &  11 &  377 &  46 & 8.1 \\
                                    & $S$=$7$, $\lambda$=$3.9$, EF & 1.97 &  12 &  363 &  45 & 8.1 \\
                                    & $S$=$8$, $\lambda$=$4.3$, EF & 1.94 &  10 &  606 &  69 & 8.8 \\
                                    & $S$=$8$, $\lambda$=$4.5$, EF & 1.92 &  10 &  577 &  66 & 8.7 \\
                                    & $S$=$9$, $\lambda$=$5.1$, EF & 1.89 &  11 & 1\,009 & 109 & 9.3 \\
                                   & $S$=$10$, $\lambda$=$6.0$, EF & 1.85 &  11 & 1\,687 & 180 & 9.4 \\
                                   & $S$=$11$, $\lambda$=$6.3$, EF & 1.85 &  10 & 3\,162 & 333 & 9.5 \\
                                   & $S$=$12$, $\lambda$=$7.2$, EF & 1.82 &  10 & 5\,322 & 553 & 9.6 \\
                                   \addlinespace[0.3em]
            \mr{\PHastPlus{} \eqref{eq:plus}}& $S$=$8$, $\lambda$=$4.15$, EF & 2.24 & 15 &  61 & 15 & 4.1 \\
                       & $S$=$10$, $\lambda$=$5.15$, EF & 2.08 & 13 &  63 & 15 & 4.2 \\
                        & $S$=$11$, $\lambda$=$5.7$, EF & 2.03 & 12 &  68 & 15 & 4.4 \\
                        & $S$=$12$, $\lambda$=$6.2$, EF & 2.02 & 12 &  80 & 18 & 4.5 \\
                        & $S$=$11$, $\lambda$=$6.5$, EF & 1.96 & 13 &  73 & 17 & 4.4 \\
                        \addlinespace[0.3em]
      \mr{\PHastPlus{} \eqref{eq:plus:wrap}}& $\delta$=$2$, $S$=$8$, $\lambda$=$4.1$, C & 2.50 & 10 &  81 & 17 & 4.7 \\
                           & $\delta$=$3$, $S$=$8$, $\lambda$=$4.15$, C & 2.37 & 11 &  90 & 19 & 4.8 \\
          & $\delta$=$1$, $S$=$8$, $\lambda$=$4.3$, $L$=$1024$, EF & 2.18 & 14 &  71 & 16 & 4.3 \\
        & $\delta$=$2$, $S$=$8$, $\lambda$=$4.35$, EF & 2.07 & 12 &  80 & 17 & 4.6 \\
         & $\delta$=$2$, $S$=$8$, $\lambda$=$4.5$, EF & 2.05 & 11 &  81 & 17 & 4.6 \\
        & $\delta$=$3$, $S$=$8$, $\lambda$=$4.35$, EF & 2.03 & 11 &  92 & 19 & 4.9 \\
         & $\delta$=$2$, $S$=$8$, $\lambda$=$4.7$, EF & 2.03 & 12 &  82 & 18 & 4.6 \\
    & $\delta$=$1$, $S$=$10$, $\lambda$=$5.55$, $L$=$2048$, EF & 1.96 & 11 &  83 & 18 & 4.7 \\
        & $\delta$=$1$, $S$=$11$, $\lambda$=$6.1$, EF & 1.93 & 11 & 105 & 21 & 5.0 \\
     & $\delta$=$1$, $S$=$10$, $\lambda$=$6.2$, $L$=$2048$, EF & 1.90 & 13 &  87 & 19 & 4.7 \\
        & $\delta$=$1$, $S$=$12$, $\lambda$=$6.7$, EF & 1.88 & 10 & 129 & 24 & 5.3 \\
        & $\delta$=$1$, $S$=$11$, $\lambda$=$6.6$, EF & 1.85 & 11 & 114 & 22 & 5.1 \\
        & $\delta$=$1$, $S$=$12$, $\lambda$=$7.1$, EF & 1.84 & 10 & 140 & 26 & 5.4 \\
            \addlinespace[0.3em]
    \mr{PtrHash}& $\lambda$=$3.0$, $\alpha$=$0.99$, linear, Vec & 2.99 &  16 &   76 &  21 & 3.6 \\
    & $\lambda$=$3.0$, $\alpha$=$0.99$, linear, cEF & 2.78 &  17 &   75 &  21 & 3.6 \\
                    & $\lambda$=$3.5$, $\alpha$=$0.99$, cubic, cEF & 2.40 &  15 &   98 &  21 & 4.7 \\
                    & $\lambda$=$4.0$, $\alpha$=$0.99$, cubic, cEF & 2.12 &  14 &  192 &  37 & 5.1 \\
                    \addlinespace[0.3em]
                               \mr{SIMDRecSplit}& $n$=$5$, $b$=$5$ & 2.96 & 142 &   60 &  15 & 3.9 \\
                                              & $n$=$8$, $b$=$100$ & 1.81 & 104 &  204 &  35 & 5.8 \\
                                              \addlinespace[0.3em]
          \mr{SicHash}& $\alpha$=$0.9$, $p_1$=$21\%$, $p_2$=$78\%$ & 2.41 &  69 &  134 &  29 & 4.6 \\
                     & $\alpha$=$0.97$, $p_1$=$45\%$, $p_2$=$31\%$ & 2.08 &  64 &  188 &  33 & 5.7 \\

    \bottomrule
\end{tabular}
\end{centering}

%% file: table-500M-extra.tex
\newcommand{\mr}[1]{\multirow{2}{*}{#1}}
\begin{centering}\small
  \begin{tabular}[t]{ll rrrrr}
      \toprule
      Method & Configuration & Space      & Query      & \multicolumn{3}{c}{Construction [ns/key]} \\
               &               & {\scriptsize [bit/key]} & {\scriptsize [ns/query]} & {\scriptsize 1 thread} & {\scriptsize 12 threads} & {\scriptsize speedup} \\ \midrule
                \mr{PHOBIC}& $\lambda$=$3.9$, $\alpha$=$1.0$, IC-C & 3.25 &  95 &  252 &  38 & 6.7 \\
                           & $\lambda$=$6.5$, $\alpha$=$1.0$, IC-C & 2.39 &  88 &  971 & 123 & 7.9 \\
                           & $\lambda$=$4.5$, $\alpha$=$1.0$, IC-R & 2.11 & 155 &  307 &  44 & 7.0 \\
                           & $\lambda$=$6.5$, $\alpha$=$1.0$, IC-R & 1.85 & 136 &  973 & 123 & 7.9 \\
                           & $\lambda$=$9.0$, $\alpha$=$1.0$, IC-R & 1.74 & 124 & 7\,576 & 920 & 8.2 \\
                           \addlinespace[0.3em]
                \mr{PTHash}& $\lambda$=$4.0$, $\alpha$=$0.99$, C-C & 3.19 &  67 &  874 & 326 & 2.7 \\
                            & $\lambda$=$5.0$, $\alpha$=$0.99$, EF & 2.11 & 135 & 1\,807 & 654 & 2.8 \\
                            \addlinespace[0.3em]
            \mr{PTHash-HEM}& $\lambda$=$4.0$, $\alpha$=$0.99$, C-C & 3.17 &  76 &  884 & 119 & 7.4 \\
                            & $\lambda$=$5.0$, $\alpha$=$0.99$, EF & 2.11 & 147 & 1\,793 & 232 & 7.7 \\
                            \addlinespace[0.3em]
                                         \mr{FMPH}& $\gamma$=$2.0$ & 3.40 & 105 &  121 &  24 & 5.0 \\
                                                  & $\gamma$=$1.0$ & 2.80 & 145 &  181 &  36 & 5.0 \\
                                                  \addlinespace[0.3em]
                        \mr{FMPHGO}& $\gamma$=$2.0, s$=$4, b$=$16$ & 2.86 &  77 &  893 & 177 & 5.0 \\
                                   & $\gamma$=$1.0, s$=$4, b$=$16$ & 2.21 & 133 & 1\,167 & 251 & 4.6 \\
                                   \addlinespace[0.3em]
                          \mr{PHast}& $S$=$4$, $\lambda$=$2.6$, EF & 2.33 &  51 &  152 &  31 & 4.9 \\
                                    & $S$=$5$, $\lambda$=$2.8$, EF & 2.18 &  39 &  189 &  33 & 5.8 \\
                          & $S$=$6$, $\lambda$=$3.2$, EF & 2.07 &  34 &  266 &  39 & 6.8 \\
                                     & $S$=$8$, $\lambda$=$4.2$, C & 2.07 &  23 &  638 &  75 & 8.5 \\
                                    & $S$=$7$, $\lambda$=$3.7$, EF & 2.00 &  30 &  397 &  52 & 7.7 \\
                                    & $S$=$7$, $\lambda$=$3.9$, EF & 1.97 &  32 &  383 &  50 & 7.6 \\
                                    & $S$=$8$, $\lambda$=$4.3$, EF & 1.94 &  25 &  623 &  73 & 8.5 \\
                                    & $S$=$8$, $\lambda$=$4.5$, EF & 1.92 &  27 &  594 &  70 & 8.5 \\
                                    & $S$=$9$, $\lambda$=$5.1$, EF & 1.89 &  30 & 1\,030 & 114 & 9.1 \\
                                   & $S$=$10$, $\lambda$=$6.0$, EF & 1.85 &  32 & 1\,708 & 181 & 9.4 \\
                                   & $S$=$11$, $\lambda$=$6.3$, EF & 1.85 &  28 & 3\,182 & 334 & 9.5 \\
                                   & $S$=$12$, $\lambda$=$7.2$, EF & 1.82 &  29 & 5\,341 & 555 & 9.6 \\
                                   \addlinespace[0.3em]
       \mr{\PHastPlus{} \eqref{eq:plus}}& $S$=$8$, $\lambda$=$4.15$, EF & 2.24 & 35 &  74 & 20 & 3.8 \\
                       & $S$=$10$, $\lambda$=$5.15$, EF & 2.08 & 34 &  81 & 20 & 4.1 \\
                        & $S$=$11$, $\lambda$=$5.7$, EF & 2.03 & 33 &  87 & 21 & 4.2 \\
                        & $S$=$12$, $\lambda$=$6.2$, EF & 2.01 & 32 &  98 & 23 & 4.4 \\
                        & $S$=$11$, $\lambda$=$6.5$, EF & 1.95 & 38 &  93 & 22 & 4.3 \\
                                   \addlinespace[0.3em]
      \mr{\PHastPlus{} \eqref{eq:plus:wrap}} & $\delta$=$2$, $S$=$8$, $\lambda$=$4.1$, C & 2.55 & 22 &  94 & 22 & 4.3 \\
      & $\delta$=$3$, $S$=$8$, $\lambda$=$4.15$, C & 2.41 & 24 & 107 & 23 & 4.6 \\
      & $\delta$=$1$, $S$=$8$, $\lambda$=$4.3$, $L$=$1024$, EF & 2.18 & 29 &  88 & 21 & 4.2 \\
        & $\delta$=$2$, $S$=$8$, $\lambda$=$4.35$, EF & 2.07 & 26 &  98 & 22 & 4.4 \\
         & $\delta$=$2$, $S$=$8$, $\lambda$=$4.5$, EF & 2.05 & 27 &  98 & 22 & 4.5 \\
        & $\delta$=$3$, $S$=$8$, $\lambda$=$4.35$, EF & 2.03 & 25 & 108 & 23 & 4.6 \\
        & $\delta$=$2$, $S$=$8$, $\lambda$=$4.7$, EF & 2.02 & 29 & 100 & 22 & 4.5 \\
    & $\delta$=$1$, $S$=$10$, $\lambda$=$5.55$, $L$=$2048$, EF & 1.95 & 28 & 102 & 23 & 4.4 \\
        & $\delta$=$1$, $S$=$11$, $\lambda$=$6.1$, EF & 1.93 & 28 & 125 & 26 & 4.8 \\
     & $\delta$=$1$, $S$=$10$, $\lambda$=$6.2$, $L$=$2048$, EF & 1.90 & 33 & 107 & 24 & 4.5 \\
        & $\delta$=$1$, $S$=$12$, $\lambda$=$6.7$, EF & 1.88 & 27 & 148 & 30 & 5.0 \\
        & $\delta$=$1$, $S$=$11$, $\lambda$=$6.6$, EF & 1.85 & 30 & 133 & 27 & 4.9 \\
        & $\delta$=$1$, $S$=$12$, $\lambda$=$7.1$, EF & 1.83 & 28 & 158 & 31 & 5.1 \\
                  \addlinespace[0.3em]
    \mr{PtrHash}& $\lambda$=$3.0$, $\alpha$=$0.99$, linear, Vec & 2.99 &  25 &   77 &  20 & 3.9 \\
    & $\lambda$=$3.0$, $\alpha$=$0.99$, linear, cEF & 2.78 &  25 &   76 &  19 & 3.9 \\
                    & $\lambda$=$3.5$, $\alpha$=$0.99$, cubic, cEF & 2.40 &  32 &  100 &  20 & 5.1 \\
                    & $\lambda$=$4.0$, $\alpha$=$0.99$, cubic, cEF & 2.12 &  31 &  194 &  39 & 5.0 \\
                    \addlinespace[0.3em]
                               \mr{SIMDRecSplit}& $n$=$5$, $b$=$5$ & 2.96 & 295 &   61 &  16 & 3.9 \\
                                              & $n$=$8$, $b$=$100$ & 1.81 & 215 &  197 &  34 & 5.8 \\
                        \addlinespace[0.3em]
          \mr{SicHash}& $\alpha$=$0.9$, $p_1$=$21\%$, $p_2$=$78\%$ & 2.41 & 149 &  137 &  30 & 4.5 \\
                     & $\alpha$=$0.97$, $p_1$=$45\%$, $p_2$=$31\%$ & 2.08 & 135 &  193 &  34 & 5.7 \\
    \bottomrule
\end{tabular}
\end{centering}

%% file: construction_steps.tex
\subsection{PHast Construction Steps.}\label{sec:benchmark:steps}


\begin{table*}
  \caption[Contribution of each step to PHast construction time]{Percentage contribution of the following consecutive steps to the PHast and \PHastPlus \eqref{eq:plus} construction time: mapping keys to hash codes / sorting hash codes / assigning seeds to buckets (in bold) / picking bumped keys for further processing. Only the construction of the first map-or-bump function (see Section~\ref{sec:mphf}), which considers all input keys, is taken into account. Subsequent ones are ignored as processing only a small fraction of the keys.}\label{tab:steps} 
  \centering
  \setlength{\tabcolsep}{9pt}
  \begin{tabular}{cccccccc}
    \toprule
    \multirow{2}{*}{Method} & \multirow{2}{*}{S} & \multirow{2}{*}{$\lambda$} & \multirow{2}{*}{Threads} & \multicolumn{2}{c}{$8$ byte integer keys} & \multicolumn{2}{c}{$\sim 30$ byte string keys} \\
                & & & & $5 \cdot 10^7$ & $5 \cdot 10^8$ & $5 \cdot 10^7$ & $5 \cdot 10^8$ \\
    \midrule
\multirow{6}{*}{PHast} & \multirow{2}{*}{$6$} & \multirow{2}{*}{$3.2$} & 1    & 1/5/\textbf{90}/5       & 0/4/\textbf{85}/11       & 3/4/\textbf{86}/7       & 3/3/\textbf{76}/17 \\
                                    & & & 12  & 5/9/\textbf{73}/12      & 4/12/\textbf{65}/19    & 9/9/\textbf{66}/17      & 8/10/\textbf{56}/26 \\
   \addlinespace[0.3em]
 & \multirow{2}{*}{$8$} & \multirow{2}{*}{$4.5$} & 1    & 0/2/\textbf{97}/1       & 0/2/\textbf{96}/2      & 1/2/\textbf{95}/2       & 1/2/\textbf{93}/4 \\
                                    & & & 12  & 2/5/\textbf{88}/5       & 2/6/\textbf{84}/8      & 5/4/\textbf{84}/7       & 4/6/\textbf{79}/11 \\
    \addlinespace[0.3em]
 & \multirow{2}{*}{$10$} & \multirow{2}{*}{$6.0$} & 1   & 0/1/\textbf{99}/0      & 0/1/\textbf{98}/1      &  0/1/\textbf{98}/1       & 0/1/\textbf{97}/2 \\
                                    & & & 12 & 1/2/\textbf{96}/2      & 1/2/\textbf{94}/3      & 2/2/\textbf{94}/3       & 2/2/\textbf{91}/5 \\
    \addlinespace[0.7em]
\multirow{4}{*}{\PHastPlus \eqref{eq:plus}} & \multirow{2}{*}{$8$} & \multirow{2}{*}{$4.5$} & 1    & 3/23/\textbf{62}/12       & 2/19/\textbf{54}/24      & 13/16/\textbf{45}/27       & 9/11/\textbf{32}/48 \\
                                    & & & 12  & 14/28/\textbf{30}/28       & 10/29/\textbf{23}/38      & 20/20/\textbf{24}/35        & 15/21/\textbf{16}/48 \\
    \addlinespace[0.3em]
 & \multirow{2}{*}{$10$} & \multirow{2}{*}{$6.0$} & 1   & 2/18/\textbf{63}/17      & 2/13/\textbf{47}/38      & 9/12/\textbf{42}/37       & 8/10/\textbf{34}/49 \\
                                    & & & 12 & 12/25/\textbf{36}/27      & 9/26/\textbf{26}/40      & 20/19/\textbf{29}/33       & 14/20/\textbf{20}/45 \\
     \bottomrule
  \end{tabular}
\end{table*}

Table \ref{tab:steps} shows the contribution 
of the following steps to the PHast and \PHastPlus construction time: mapping keys to hash codes, sorting hash codes, assigning seeds to buckets, picking bumped keys for further processing.

As expected, the construction time of regular PHast is dominated by seed assignment, which is greatly accelerated in \PHastPlus{}.
Simultaneously, the contribution of this step is reduced by the following factors:
\begin{itemize}
  \item larger keys, which are slower to hash and pick (so the share of the first and last steps increases);
  \item multithreading, which confirms that seed assignment is efficiently parallelized (better than other steps);
  \item greater number of keys, which confirms that the seed assignment algorithm is (more than other steps) CPU-cache friendly.
\end{itemize}

%% file: ablation_study.tex
\subsection{Ablation Study.}\label{sec:ablation}

Table \ref{tab:ablation} shows how disabling or changing individual PHast ingredients affects its size. 
It only presents sizes obtained for $p$ defined in Section~\eqref{sec:implementation}, $S=8$, $\lambda=4.5$, $5 \cdot 10^7$ uniformly random $64$-bit integer keys, but our qualitative conclusions do not depend on the parameters used.

Examining all the seeds in a bucket to heuristically select the best one is time-consuming. Therefore, simply assigning the first feasible seed may seem like an attractive way to significantly reduce construction time.
However, the increase in size (from $1.92$ to $2.39$ bits/key) is too large to make this worthwhile. It is better to use a smaller seed size instead. Even $S=4$ can yield $2.32$ bits/key (see Figure \ref{fig:optimal:lambda}).

\begin{table*}[tb]
  \caption{Sizes of different PHast $S=8$, $\lambda=4.5$ variants, measured for $5 \cdot 10^7$ keys. For each variant, we give two sizes: with the usual slice restriction on key values ($L=1024$) and without it ($L=\infty$; each key can be mapped to any value).}\label{tab:ablation}
  \centering
  \begin{tabular}{lcc}
    \toprule
    PHast variant (differences from the ordinary version) & \multicolumn{2}{c}{size [bit/key]} \\
    \multicolumn{1}{r}{$L=$} & $1024$ & $\infty$ \\
    \midrule
    ordinary & $1.92$ & $3.47$ \\
    assigning the first feasible seed to the buckets (without heuristics) & $2.39$ & $3.46$ \\
    $W=1$ (processes buckets in ascending index order, ignoring their sizes) & $2.20$ & $3.47$ \\
    $W=\infty$ (sorting all buckets in advance), standard bucket priority & $1.92$ & $3.47$ \\
    $W=\infty$, processing buckets from the largest, in ascending index order at ties & $2.36$ & $2.43$ \\ 
    \hspace*{1em} as above + assigning the first feasible seed to the buckets & $2.40$ & $2.43$ \\
    $W=\infty$, processing buckets from the largest, ties resolved randomly & $2.44$ & $2.43$  \\
    \hspace*{1em} as above + assigning the first feasible seed to the buckets & $2.44$  & $2.43$ \\
    \bottomrule
  \end{tabular}
\end{table*}

For the same reason, it is not worth reducing the window size to $W=1$ to avoid priority queue operations (especially since they consume a negligible portion of time). On the other hand, increasing the window above $W=256$, even to accommodate all the buckets ($W=\infty$), brings no benefit and is just an unnecessary waste of resources.
The function assigning priorities to buckets in the window causes them to be processed in roughly ascending index order anyway.
At the same time, replacing this function with one that considers only (or mainly) bucket sizes usually increases the resulting PHF size.
This is largely because processing buckets in ascending index order works in tandem with selecting seeds minimizing the sum of values.
When instead the first feasible seeds are assigned to buckets, taking the bucket index into account no longer significantly affects the size, still not increasing it though.

However, the situation is different when each key can be mapped to any value, without slice restrictions.
Then it is best to process buckets in descending size order, ignoring their indexes, as PtHash or CHD do.
This leads to a size ($2.43$ bits/key) that is larger than that offered by PHast, as well as by PtHash or CHD.
However, this size is still attractive for an algorithm that, by using a fixed-size seeds, could offer faster evaluation than CHD and perhaps even PtHash variants of similar size.
PHast could therefore be competitive with these methods even using only its bumping mechanism that permits fixed-size seeds, without other improvements. 

%% file: external_memory.tex
\section{External Memory Construction.}\label{ss:external}

An external memory variant of PHast construction is relatively simple and similar to many other
PHFs. We describe it anyway as applications where the PHF fits in main memory but not the key set
are an important use case of PHFs. We describe a sequential algorithm but parallelization is an orthogonal issue (see Section~\ref{sec:parallel}) that can be combined with the external algorithm described here.

Assuming the keys are stored in a file $F$, $F$ is scanned, and their hash codes (HCs) are
fed into a pipelined \cite{DKS08} external memory integer sorting algorithm, sorting by bucket ID.
The $W$ most recent buckets of the output form the current window for the map-or-bump seed search algorithm which outputs seeds and stores them in internal memory. Next, $F$ is scanned again to identify bumped keys. This needs a random access to the seed array for each key. Bumped keys are fed into the next layer of the construction algorithm. This and all subsequent layers of the PHast data structure can be computed internally in practice. Overall, external PHast construction requires I/O-volume for scanning the input twice plus pipelined sorting of $n$ $u$-bit integers which requires $2nu$ bits of $I/O$ volume in practice (for example using multiway mergesort), $un$ bits for writing sorted runs, and $un$ bits for reading the runs again before outputting a sorted data stream.

As a theoretical exercise, we can outline a fully external algorithm that only needs internal memory of size $O(W\lambda)$ beyond the one needed for external sorting. That one could keep input positions with the HCs. For identifying bumped keys their indices would be stored and sorted. In the second pass through $F$ we would then only use the bumped keys. An nice side effect is that this means we need only close to $n$ evaluations of the hash function.

%% file: advanced.tex
\section{Advanced Construction Algorithms.}\label{s:advanced}

The PHast construction algorithm we currently use is quite simple, processing buckets
in an order only dependent on their size and position.
We also tried more sophisticated algorithms that take interdependencies between buckets into account. There is evidence that with the right algorithm or sufficient
combinatorial search we can get much closer to the lower bound.
However, so far our attempts were not very successful.
At fixed seed size $S$, they save only a marginal amount of space (up to $0.08$ bits per key for $S=8$) but cost at least an order of magnitude in construction time.
Indeed, it is more economic to increase seed size. We still report on these algorithms as they might help to guide the search
for better approaches. 

\paragraph*{Smallest number of choices first:}
Rather than processing buckets in fixed order, it is slightly better to
always assign the seed of the bucket with smallest number of remaining feasible choices.
As the data structure for maintaining the required information has some overhead,
one can refine that to only look at the buckets in the current window of size $W$.
A promising observation we made is that with this heuristics, failed assignments are 
often preceded by assignments with very few open choices. This may open the way
for a backtracking algorithm changing a few previous decisions to avoid the failure.
It may be interesting to exploit that for future work.

\paragraph*{Do little harm:}
Currently we decide on the the seed that minimizes the sum of the function values.
One can also take into account the feasible choices of other buckets that are destroyed
by a seed selection. We found out that a good way to do this is to scale this harm
by the number of open choices available to the harmed bucket.
Variants of this%
\footnote{For example, let $c_i$ denote the number of remaining choices for bucket $i$; with $S=8$, $\lambda=4.6$, and a cost function $10\,000 c_i+\sum_{j\in\mathrm{harmedBuckets}(i)}5000/c_j$, we obtain $1.85$ bits per key with construction throughput of about 12\,000 keys per second.}
yielded the lowest space we have observed so far. However the
involved bookeeping is very expensive.

\paragraph*{Peeling:}
One of the original inspirations for PHast were static retrieval data structures \cite{Walzer21}
based on \emph{hypergraph peeling}. Rather than greedily placing keys, peeling greedily fills empty slots once there is only one remaining way to fill them.
This turns out to be optimal in certain situations.
We made a few attempts to adapt peeling to bucket placement.
Unfortunately this is complicated as filling one slot with a key from a bucket can have ``collateral damage'' by placing other keys elsewhere, thus harming other unseeded buckets. Overall, these attempts were not successful so far. However, we believe that
using a peeling algorithm to place size-one buckets may help.

\paragraph*{Multiple iterations with boosting:}
We tried to solve seed assignments in an iterative way based on the do-little-harm approach explained above. We recorded which buckets failed in previous iterations and
increased the penalty for harming them accordingly. The hope was that this would prioritize hard-to-place buckets sufficiently to improve the overall assignment.
However, this did not work at all. Apparently, boosting one bucket harms so many other buckets that the overall effect is negative.

\paragraph*{Local search:} When no seeds can be found for a bucket, one can
enforce a successful placement by undoing the placement of other buckets.
This has been tried in PHOBIC \cite{Hermann2023_1000164413}
but it was concluded that this was too slow when trying to achieve
high space-efficiency.
Local search is successfully used in PtrHash \cite{grootkoerkamp:LIPIcs.SEA.2025.21} but likely this is a reason why PtrHash needs more space than PHOBIC or PHast -- apparently local search only converges with enough empty cells and sufficiently small buckets. One could also integrate local search into PHast, obtaining a hybrid between PHast and PtrHash. This could be interesting if one wants to get rid of bumping to achieve better worst case query time.

%% file: conclusions.tex
\section{Conclusions and Future Work.}\label{sec:conclusions}

PHast is about as fast as it gets with respect to query time of perfect hashing.
Compared to the most simple conceivable PHF based on bucket placement,
the only apparent ``overheads'' implied by the query specified by Equation~\eqref{eq:query}
are the linear offset term $\lfloor \alpha h(k) \rfloor$ and the bumping test.
However, the offset massively reduces the range of $p$ and
allowing bumping almost halves the number of bits needed for the hash function $h$.

For a highly engineered data structure, PHast is also surprisingly simple because many of the improvements of PHast are about \emph{dropping} things previously thought necessary, for example variable bit length encoding, nonlinear bucket assignment, injective hash functions, or parallelization by explicit partitioning.  Indeed, several of our own ambitions for further sophistication (described in Section~\ref{s:advanced}) turned out to be of little value.

We view it as likely, that the space consumption is similar to PHOBIC and other bucket-placement PHFs with space overhead
$O(\log(\lambda)/\lambda)$ for expected bucket size $\lambda$, that is,
we can achieve arbitrarily small overhead in principle.
However, proving this is future work.
Another interesting loose end is whether some kind of more clever search can reduce the space overhead further.

Other interesting future work is implementing the external construction outlined in Section~\ref{ss:external}, supporting batch queries similar to
PtrHash, and using GPUs.

PHast is straightforward to generalize for $k$-perfect hashing where up to $k$ keys are allowed to map to the same value.

%% file: phast.bbl
\begin{thebibliography}{10}

\bibitem{BelazzouguiBD09}
{\sc D.~Belazzougui, F.~C. Botelho, and M.~Dietzfelbinger}, {\em Hash,
  displace, and compress}, in 17th European Symposium on Algorithms (ESA),
  vol.~5757 of LNCS, 2009, pp.~682--693.

\bibitem{BSuccinctGit}
{\sc P.~Beling}, {\em Bsuccinct}.
\newblock \url{https://github.com/beling/bsuccinct-rs} [accessed 4 Apr 2025].

\bibitem{beling2023fingerprinting}
{\sc P.~Beling}, {\em Fingerprinting-based minimal perfect hashing revisited},
  {ACM} J. Exp. Algorithmics, 28 (2023), pp.~1.4:1--1.4:16,
  \url{https://doi.org/10.1145/3596453}.

\bibitem{beling2024BSuccinct}
{\sc P.~Beling}, {\em Bsuccinct: Rust libraries and programs focused on
  succinct data structures}, SoftwareX, 26 (2024), p.~101681,
  \url{https://doi.org/https://doi.org/10.1016/j.softx.2024.101681},
  \url{https://www.sciencedirect.com/science/article/pii/S2352711024000529}.

\bibitem{FigShare}
{\sc P.~Beling, H.-P. Lehmann, and S.~Hermann}, {\em {MPHF-Experiments (fork
  with PHast+ support)}},  (2025),
  \url{https://doi.org/10.6084/m9.figshare.30272380.v6},
  \url{https://figshare.com/articles/software/MPHF-Experiments_fork_with_PHast_support_/30272380}.

\bibitem{bez2023high}
{\sc D.~Bez, F.~Kurpicz, H.-P. Lehmann, and P.~Sanders}, {\em High performance
  construction of {RecSplit} based minimal perfect hash functions}, in {ESA},
  vol.~274 of LIPIcs, Schloss Dagstuhl - Leibniz-Zentrum f{\"{u}}r Informatik,
  2023, pp.~19:1--19:16, \url{https://doi.org/10.4230/LIPICS.ESA.2023.19}.

\bibitem{botelho2007simple}
{\sc F.~C. Botelho, R.~Pagh, and N.~Ziviani}, {\em Simple and space-efficient
  minimal perfect hash functions}, in {WADS}, vol.~4619 of Lecture Notes in
  Computer Science, Springer, 2007, pp.~139--150,
  \url{https://doi.org/10.1007/978-3-540-73951-7_13}.

\bibitem{chapman2011meraculous}
{\sc J.~A. Chapman, I.~Ho, S.~Sunkara, S.~Luo, G.~P. Schroth, and D.~S.
  Rokhsar}, {\em Meraculous: De novo genome assembly with short paired-end
  reads}, PLOS ONE, 6 (2011), pp.~1--13,
  \url{https://doi.org/10.1371/journal.pone.0023501},
  \url{https://doi.org/10.1371/journal.pone.0023501}.

\bibitem{chazelle2004bloomier}
{\sc B.~Chazelle, J.~Kilian, R.~Rubinfeld, and A.~Tal}, {\em The bloomier
  filter: an efficient data structure for static support lookup tables}, in
  {SODA}, {SIAM}, 2004, pp.~30--39.

\bibitem{DKS08}
{\sc R.~Dementiev, L.~Kettner, and P.~Sanders}, {\em {STXXL}: {S}tandard
  {T}emplate {L}ibrary for {XXL} data sets}, Software Practice {\&} Experience,
  38 (2008), pp.~589--637.

\bibitem{fxhash}
{\sc T.~R.~P. Developers}, {\em Fxhash}.
\newblock \url{https://github.com/cbreeden/fxhash} [accessed 4 Apr 2025].

\bibitem{dillinger2022burr}
{\sc P.~C. Dillinger, L.~H{\"{u}}bschle{-}Schneider, P.~Sanders, and
  S.~Walzer}, {\em Fast succinct retrieval and approximate membership using
  ribbon}, in {SEA}, vol.~233 of LIPIcs, Schloss Dagstuhl - Leibniz-Zentrum
  f{\"{u}}r Informatik, 2022, pp.~4:1--4:20,
  \url{https://doi.org/10.4230/LIPICS.SEA.2022.4}.

\bibitem{DHSW22}
{\sc P.~C. Dillinger, L.~H{\"{u}}bschle{-}Schneider, P.~Sanders, and
  S.~Walzer}, {\em Fast succinct retrieval and approximate membership using
  ribbon}, in 20th Symposium on Experimental Algorithms ({SEA}), vol.~233 of
  LIPIcs, 2022, pp.~4:1--4:20.
\newblock best paper award.

\bibitem{Elias74}
{\sc P.~Elias}, {\em Efficient storage and retireval by content and adress of
  static files}, Journal of the ACM, 21 (1974), pp.~246--260.

\bibitem{esposito2020recsplit}
{\sc E.~Esposito, T.~M. Graf, and S.~Vigna}, {\em {RecSplit}: Minimal perfect
  hashing via recursive splitting}, in {ALENEX}, {SIAM}, 2020, pp.~175--185,
  \url{https://doi.org/10.1137/1.9781611976007.14}.

\bibitem{Fano71}
{\sc R.~M. Fano}, {\em On the number of bits required to implement an
  associative memory}, tech. report, MIT, Computer Structures Group, 1971.
\newblock Project MAC, Memorandum 61.

\bibitem{fox1992faster}
{\sc E.~A. Fox, Q.~F. Chen, and L.~S. Heath}, {\em A faster algorithm for
  constructing minimal perfect hash functions}, in {SIGIR}, {ACM}, 1992,
  pp.~266--273, \url{https://doi.org/10.1145/133160.133209}.

\bibitem{MPHFLowerBound1984}
{\sc M.~L. Fredman and J.~Komlós}, {\em On the size of separating systems and
  families of perfect hash functions}, SIAM Journal on Algebraic Discrete
  Methods, 5 (1984), pp.~61--68, \url{https://doi.org/10.1137/0605009}.

\bibitem{genuzio2016fast}
{\sc M.~Genuzio, G.~Ottaviano, and S.~Vigna}, {\em Fast scalable construction
  of (minimal perfect hash) functions}, in {SEA}, vol.~9685 of Lecture Notes in
  Computer Science, Springer, 2016, pp.~339--352,
  \url{https://doi.org/10.1007/978-3-319-38851-9_23}.

\bibitem{gxhash}
{\sc O.~Giniaux}, {\em Gxhash: A high-throughput, non-cryptographic hashing
  algorithm leveraging modern cpu capabilities}, Sept. 2023,
  \url{https://doi.org/10.5281/zenodo.8368254},
  \url{https://doi.org/10.5281/zenodo.8368254}.

\bibitem{grootkoerkamp:LIPIcs.SEA.2025.21}
{\sc R.~Groot~Koerkamp}, {\em {PtrHash: Minimal Perfect Hashing at RAM
  Throughput}}, in 23rd International Symposium on Experimental Algorithms
  (SEA), P.~Mutzel and N.~Prezza, eds., vol.~338 of (LIPIcs), Dagstuhl,
  Germany, 2025, Schloss Dagstuhl -- Leibniz-Zentrum f{\"u}r Informatik,
  pp.~21:1--21:21, \url{https://doi.org/10.4230/LIPIcs.SEA.2025.21},
  \url{https://drops.dagstuhl.de/entities/document/10.4230/LIPIcs.SEA.2025.21}.

\bibitem{Hermann2023_1000164413}
{\sc S.~Hermann}, {\em Accelerating minimal perfect hash function construction
  using gpu parallelization}, master's thesis, Karlsruher Institut für
  Technologie (KIT), 2023, \url{https://doi.org/10.5445/IR/1000164413}.

\bibitem{Lehmann:MPHFExperiments}
{\sc H.~Lehmann}, {\em Mphf{-}experiments}.
\newblock \url{https://github.com/ByteHamster/MPHF-Experiments} [accessed 4 Apr
  2025].

\bibitem{Lehmann24a}
{\sc H.~Lehmann}, {\em Fast and Space-Efficient Perfect Hashing}, PhD thesis,
  Karlsruhe Institute of Technology, Germany, 2024,
  \url{https://nbn-resolving.org/urn:nbn:de:101:1-2412040400410.467433526210}.

\bibitem{LSW23slick}
{\sc H.~Lehmann, P.~Sanders, and S.~Walzer}, {\em Sliding block hashing (slick)
  -- basic algorithmic ideas}, CoRR, abs/2304.09283 (2023),
  \url{https://doi.org/10.48550/arXiv.2304.09283},
  \url{https://doi.org/10.48550/arXiv.2304.09283},
  \url{https://arxiv.org/abs/2304.09283}.

\bibitem{lehmann2025modernminimalperfecthashing}
{\sc H.-P. Lehmann, T.~Mueller, R.~Pagh, G.~E. Pibiri, P.~Sanders, S.~Vigna,
  and S.~Walzer}, {\em Modern minimal perfect hashing: A survey}, 2025,
  \url{https://arxiv.org/abs/2506.06536},
  \url{https://arxiv.org/abs/2506.06536}.

\bibitem{lehmann2023sichash}
{\sc H.-P. Lehmann, P.~Sanders, and S.~Walzer}, {\em {SicHash} -- small
  irregular cuckoo tables for perfect hashing}, in {ALENEX}, {SIAM}, 2023,
  pp.~176--189, \url{https://doi.org/10.1137/1.9781611977561.CH15}.

\bibitem{lehmann2023bipartite}
{\sc H.-P. Lehmann, P.~Sanders, and S.~Walzer}, {\em {ShockHash}: Near
  optimal-space minimal perfect hashing beyond brute-force}, arXiv preprint,
  invited to Algorithmica,  (2024),
  \url{https://doi.org/10.48550/ARXIV.2310.14959}.

\bibitem{lehmann2023shockhash}
{\sc H.-P. Lehmann, P.~Sanders, and S.~Walzer}, {\em {ShockHash}: Towards
  optimal-space minimal perfect hashing beyond brute-force}, in {ALENEX},
  {SIAM}, 2024, pp.~194--206, \url{https://doi.org/10.1137/1.9781611977929.15}.

\bibitem{lehmann2025consensus}
{\sc H.-P. Lehmann, P.~Sanders, S.~Walzer, and J.~Ziegler}, {\em {Combined
  Search and Encoding for Seeds, with an Application to Minimal Perfect
  Hashing}}, 351 (2025), pp.~109:1--109:18,
  \url{https://doi.org/10.4230/LIPIcs.ESA.2025.109},
  \url{https://drops.dagstuhl.de/entities/document/10.4230/LIPIcs.ESA.2025.109}.

\bibitem{fastrange}
{\sc D.~Lemire}, {\em Fast random integer generation in an interval}, ACM
  Trans. Model. Comput. Simul., 29 (2019),
  \url{https://doi.org/10.1145/3230636}, \url{https://doi.org/10.1145/3230636}.

\bibitem{limasset2017fast}
{\sc A.~Limasset, G.~Rizk, R.~Chikhi, and P.~Peterlongo}, {\em Fast and
  scalable minimal perfect hashing for massive key sets}, in {SEA}, vol.~75 of
  LIPIcs, Schloss Dagstuhl - Leibniz-Zentrum f{\"{u}}r Informatik, 2017,
  pp.~25:1--25:16, \url{https://doi.org/10.4230/LIPICS.SEA.2017.25}.

\bibitem{muller2014retrieval}
{\sc I.~Müller, P.~Sanders, R.~Schulze, and W.~Zhou}, {\em Retrieval and
  perfect hashing using fingerprinting}, in {SEA}, vol.~8504 of Lecture Notes
  in Computer Science, Springer, 2014, pp.~138--149,
  \url{https://doi.org/10.1007/978-3-319-07959-2_12}.

\bibitem{NelderMead1965}
{\sc J.~A. Nelder and R.~Mead}, {\em A simplex method for function
  minimization}, The Computer Journal, 7 (1965), pp.~308--313,
  \url{https://doi.org/10.1093/comjnl/7.4.308},
  \url{https://doi.org/10.1093/comjnl/7.4.308},
  \url{https://arxiv.org/abs/https://academic.oup.com/comjnl/article-pdf/7/4/308/1013182/7-4-308.pdf}.

\bibitem{pibiri2021pthash}
{\sc G.~E. Pibiri and R.~Trani}, {\em {PTHash}: Revisiting {FCH} minimal
  perfect hashing}, in 44th {ACM} Conference on Research and Development in
  Information Retrieval ({SIGIR}), {ACM}, 2021, pp.~1339--1348,
  \url{https://doi.org/10.1145/3404835.3462849}.

\bibitem{MPHFLowerBound1992}
{\sc J.~Radhakrishnan}, {\em Improved bounds for covering complete uniform
  hypergraphs}, Information Processing Letters, 41 (1992), pp.~203 -- 207,
  \url{https://doi.org/https://doi.org/10.1016/0020-0190(92)90181-T},
  \url{http://www.sciencedirect.com/science/article/pii/002001909290181T}.

\bibitem{HLPSW24}
{\sc H.-P.~L. Stefan~Hermann, G.~Pibiri, P.~Sanders, and S.~Walzer}, {\em
  {PHOBIC}: Perfect hashing with optimized bucket sizes and interleaved
  coding}, in 32nd European Symposium on Algorithms, vol.~308 of LIPIcs, 2024.

\bibitem{Walzer21}
{\sc S.~Walzer}, {\em Peeling close to the orientability threshold -- spatial
  coupling in hashing-based data structures}, in {ACM-SIAM} Symposium on
  Discrete Algorithms (SODA), {SIAM}, 2021, pp.~2194--2211.

\end{thebibliography}
